\documentclass[fleqn,usenatbib]{mnras}

\usepackage{newtxtext,newtxmath}
\usepackage[T1]{fontenc}

\DeclareRobustCommand{\VAN}[3]{#2}
\let\VANthebibliography\thebibliography
\def\thebibliography{\DeclareRobustCommand{\VAN}[3]{##3}\VANthebibliography}

\usepackage{graphicx}	% Including figure files
\usepackage{amsmath}	% Advanced maths commands
\usepackage{supertabular}

\title{Study of Bright Compact Radio Sources of the Northern Hemisphere at the frequency of 111 MHz}

\author[S. A. Tyul'bashev et al.]{
S. A. Tyul'bashev,$^{1}$\thanks{E-mail: serg@prao.ru (SAT)}
I. V. Chashei, $^{1}$
I. A. Subaev$^{1}$
M. A. Kitaeva$^{1}$
\\
$^{1}$ P.N. Lebedev Physical Institute of the Russian Academy of Sciences, Astro Space Center, Pushchino Radio Astronomy Observatory,\\
Radiotelescopnaya 1a, Moscow reg., Pushchino, 142290, Russia \\
}

\date{January 24, 2020}

\pubyear{January 24, 2020}

\begin{document}
\label{firstpage}
\pagerange{\pageref{firstpage}--\pageref{lastpage}}
\maketitle

\begin{abstract}
The search for compact components of strong ($S_{int} \ge 5$ Jy at 102.5 MHz) discrete radio sources from the Pushchino catalogue was carried out using the method of interplanetary scintillation. A total of 3620 sources were examined, and 812 of them were found to compact (scintillating) components. Estimates of fluctuations of the flux density of these compact components were derived from the scintillation index ($m_{max}$) corresponding to an elongation of $25^{\circ}$. The angular size and compactness of 178 sources with compact components were estimated. Scintillation indices of sources corresponding to the compact component ($m_{max}$) and flux densities of compact components were determined. It was demonstrated that slow variations of the spatial distribution of interplanetary plasma, which are related to the 11-year cycle of solar activity, may exert a systematic influence on the estimates of angular sizes of sources. Coefficients compensating the deviation from the spherical symmetry of solar wind in the estimates of angular sizes were found using the coefficient of asymmetry of the statistical distribution of intensity fluctuations. The study of correlations between the parameters of sources in the sample revealed that the maximum value of the scintillation index decreases as the integrated flux increases, while the angular size has no marked dependence on the integrated flux.

\end{abstract}


%%%%%%%%%%%%%%%%%%%%%%%%%%%%%%%%%%%%%%%%%%%%%%%%%%

%%%%%%%%%%%%%%%%% BODY OF PAPER %%%%%%%%%%%%%%%%%%

\section{Introduction}
Two methods are used to estimate the angular size of compact radio sources in the meter wavelength range. The first is radio interferometric observations. If several radio telescopes are available for observations, aperture synthesis may be performed. Following data processing, a map of the studied source with identifiable compact components is obtained. If the angular size of a compact radio source is less than 1$^{\prime\prime}$, the baseline (distance between the telescopes) should be on the order of 1000 km. The second approach is the method of interplanetary scintillation. If a compact radio source is less than several angular seconds in size, its scintillation induced by moving inhomogeneities of the solar wind plasma is observed. The size of this source may then be estimated by analyzing the observed flux density fluctuations. If compact components of a source are less than several dozen milliseconds in size, both interplanetary and interstellar scintillations should be observed.

Various methods for estimating the angular size of a radio source based on the observed flux density fluctuations are known. The following observational data may be used in such studies: elongation dependence of the scintillation index in the region of weak scintillation (\citeauthor{Marians1975}, \citeyear{Marians1975}), scintillation power spectrum (\citeauthor{Shishov1978} (\citeyear{Shishov1978}), \citeauthor{Glyantsev2013} (\citeyear{Glyantsev2013})), time spectra of diffraction scintillation in the saturation region (\citeauthor{Glyantsev2013}, \citeyear{Glyantsev2013}), and coefficient of asymmetry of the statistical distribution of fluctuations of the radiation flux density (\citeauthor{Bourgois1972} (\citeyear{Bourgois1972}), \citeauthor{Shishov2005} (\citeyear{Shishov2005})).

The method of estimation of the angular size and compactness based on the asymmetry coefficient was developed in \citeauthor{Tyulbashev2019} (\citeyear{Tyulbashev2019}), where a sample of 53 strong scintillating radio sources was selected from the dataset compiled following a survey of scintillating radio sources (\citeauthor{Purvis1987}, \citeyear{Purvis1987}).

In this study, the method of estimation of the angular size and compactness of radio sources based on the asymmetry coefficient is applied to a selection of discrete sources observed with the Big Scanning Array of the Lebedev Physical Institute (BSA LPI) in a survey of the northern sky at 102.5 MHz \citeauthor{Dagkesamanskii2000} (\citeyear{Dagkesamanskii2000}).

\section{MONITORING OBSERVATIONS AT THE BSA LPI RADIO TELESCOPE AND SELECTION OF COMPACT RADIO SOURCES}

Following an extensive reconstruction that ended in 2012, the effective area of the meridional BSA LPI radio telescope increased by a factor of 2–3. Several independent beam-forming systems have been commissioned later (starting from 2013). Specifically, a system of 128 uncontrolled beams with fixed declination coordinate positioning is used for round-the-clock monitoring. A total of 96 beams from this group, which cover declinations from $-9^{\circ}$ to $+42^{\circ}$ and have an instantaneous field of view of ~50 $deg^{2}$ (if the directional pattern of each beam at half power is taken into account) are connected to digital recorders. Round-the-clock monitoring observations at the BSA LPI radio telescope are performed at a center frequency of 110.3 MHz in a 2.5-MHz-wide band. The sampling rate is ~100 ms, which allows one to detect the interplanetary scintillation of compact sources. These monitoring observations have been performed since 2014 and have insignificant time gaps due to emergency outages and routine maintenance. Parameters of the reconstructed BSA LPI radio telescope and 96-beam observation programs were detailed in \citeauthor{Shishov2016} (\citeyear{Shishov2016}), \citeauthor{Tyulbashev2016} (\citeyear{Tyulbashev2016}).

Data from five full years of round-the-clock observations of a region ~17000 $deg^{2}$ in size have already been accumulated. Earlier BSA LPI observations revealed that the rate of detection of scintillating radio sources in a survey of compact sources is one per ~1 $deg^{2}$ (\citeauthor{Artyukh1996}, \citeyear{Artyukh1996}). Therefore, the expected number of observable compact sources is ~15000–20000. However, only some of them lend themselves to the determination of the angular size. Specifically, if the asymmetry coefficient is used to estimate the angular size, one needs to know the integrated flux density of the source. This flux density should be above the confusion level for extended (nonscintillating) radio sources, which is ~0.6 Jy for BSA LPI observations at 102.5 MHz (\citeauthor{Dagkesamanskii2000}, \citeyear{Dagkesamanskii2000}). The flux density of sources suitable for estimation of the angular size should be at least several times higher than this value.

Sources with an integrated flux density above 5 Jy were selected for subsequent analysis from the catalogue of discrete radio sources (\citeauthor{Dagkesamanskii2000}, \citeyear{Dagkesamanskii2000}). This flux-density limit was chosen due to the fact that the direction of BSA LPI beams on the sky is fixed, and the declination coordinates of a source may be misaligned with the BSA beam direction. Therefore, only a fraction of the full flux of a discrete source is recorded. With the limit set to 5 Jy, the observed flux density is expected to be 3–4 times higher than the confusion level even if the declination coordinates of the studied source fall between the directions of two neighboring beams. The same limit 5 Jy was identified in \citeauthor{Dagkesamanskii2000} (\citeyear{Dagkesamanskii2000}) as a flux density that ensures completeness of the survey.

A total of 6487 sources with declinations $-9^{\circ} < \delta <+42^{\circ}$ are found in the online version1 of the Pushchino catalogue of discrete radio sources. 3620 of them have flux densities above 5 Jy. The sources with compact components were to be identified in this group for further analysis. As was demonstrated in \citeauthor{Artyukh1996a} (\citeyear{Artyukh1996a}), the weakest compact sources detected in BSA LPI observations with the interplanetary scintillation method have signal-to-noise ratio $S/N = 0.5$ at a time constant of 0.5 s. Since the processed record includes ~200 independent points taken at the maximum of the directional pattern, the resulting signal-to-noise ratio is $S/N = 200^{1/2} \times 0.5 = 7$. It was also demonstrated in \citeauthor{Artyukh1996a} (\citeyear{Artyukh1996a}) that approximately one half of the sources with $S/N = 0.5$ is left undetected in a blind search. If $S/N = 1$ is set as a criterion for selection of scintillating radio sources, only 1 source out of 1000 will possibly be left undetected. The $S/N \ge 1$ criterion was adopted for selecting scintillating radio sources from the obtained sample of sources. This corresponds to $S/N \approx 14$ for the weakest observed compact radio sources after processing of their scintillations.

The presence of a compact radio source in a selected discrete object was verified in the following way.

1) Coordinates of the source under examination were recalculated to the year of observations, and the numbers of beams between which the source is located were determined.

2) Optimum elongations of observations closest to $\epsilon =25^{\circ}$ (at 111 MHz, elongations $23^{\circ}$ – $25^{\circ}$ correspond to the maximum observed fluctuations of the flux density).

3) Raw data for days with the optimum elongation were selected, and the recorded signal was convolved with the modeled shape of the directional pattern. Since ionospheric scintillations may produce a considerable contribution in the in the meter wavelength range and shift the catalogue coordinate by up to a minute, the procedure of determination of the accurate source coordinate at the day of observations is mandatory.

4) The full length of the source record made by a diffraction-grating antenna varies with its declination as $1/\cos(\delta)$. Therefore, the expected positioning of zeroes (lack of signal) of the source traversing the directional pattern may be determined based on the source coordinate.

5) The mean-square deviation of noise was determined within an interval of 60 s (at the source top) and 30 s (at zeroes of the directional pattern). If the estimated deviation at the top was $2^{1/2}$ times (corresponds to $S/N = 1$) higher than that at zeroes of the directional pattern, the source was selected for further study.

6) All the selected sources were checked visually. This allowed us to identify those sources that had the noise track at the top expanded by interference. A fraction of selected sources were discarded after the visual inspection.

\begin{figure*}
\begin{center}
	\includegraphics[width=0.6\textwidth]{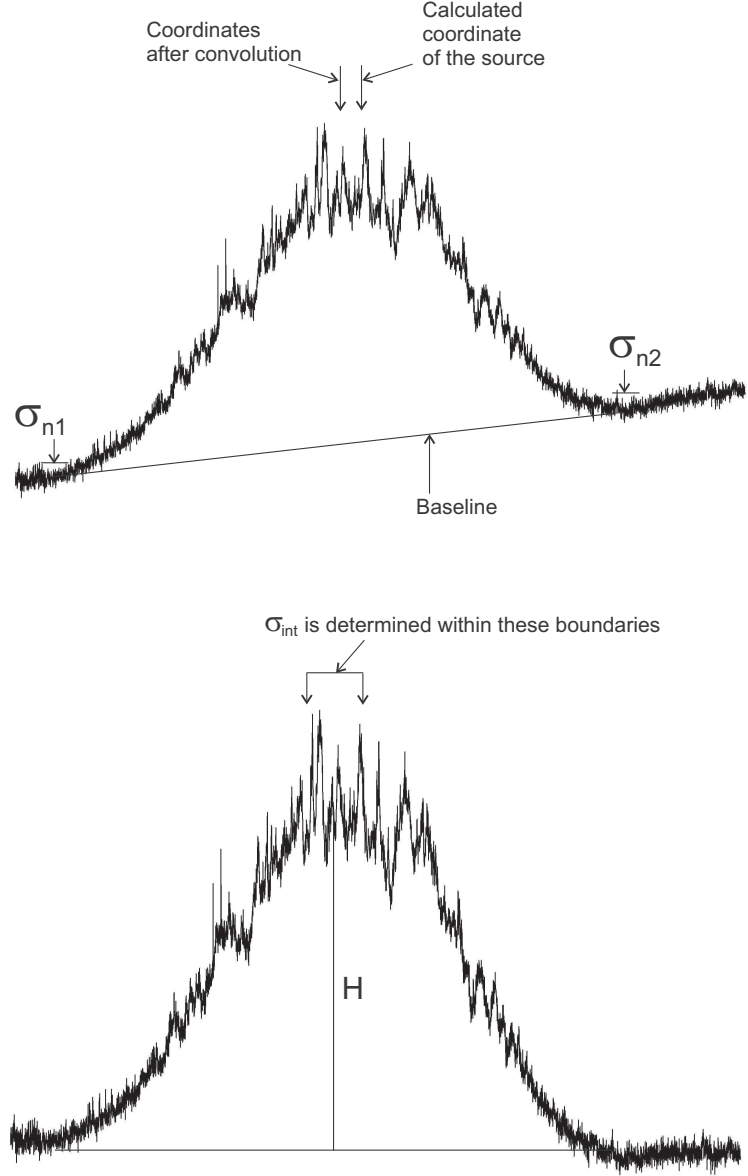}
    \caption{ The initial record of intensity of radio source 4C+21.09 is sown in the upper panel. Arrows denote the minimum signal levels in meridian passage (zeroes of the directional pattern). A straight baseline, which reflects the temperature of the Galactic background, goes through the zeroes of the pattern. Two arrows at the top denote the source coordinate calculated for the date of observations and the source coordinate obtained after convolution. The shift of the source coordinate due to ionospheric scintillations is determined by convolving the observed intensity dataset with the theoretical BSA LPI directional pattern. The same source with the baseline counts subtracted is shown in the lower panel. Short (30-second) line segments near each zero of the directional pattern indicate the boundaries within which the mean-square deviations of noise ($\sigma_{n1}$ and $\sigma_{n2}$) were determined. Vertical line $H$ indicates the radiation flux density of the source in arbitrary units, and arrows at the top denote the boundaries within which the root-mean-square noise value at the top was estimated: $\sigma_{int} = (\sigma^2_{sc} + \sigma^2_{2n} )^{1/2}$ $(\sigma^2_n =(\sigma^2_{n1} + \sigma^2_{n2}) / 2$; $\sigma^2_{sc}$; is the dispersion of flux-density fluctuations of the compact component and $\sigma^2_n$ is the dispersion of the noise signal).}
    \label{fig:fig1_TChSK}
\end{center}
\end{figure*}

Fig.~\ref{fig:fig1_TChSK} illustrates the selection of compact sources for further study.

A total of 812 sources were thus selected.

\section{ESTIMATES OF THE ANGULAR SIZES OF RADIO SOURCES IN 2014-2018}

The scintillation index and the coefficient of asymmetry of the statistical distribution of flux-density for a given observational session was estimated based on the observed intensity of the source:

%1
\begin{equation}
m^2 = \langle (I - \langle I \rangle )^2 \rangle / \langle I\rangle ^2,
\end{equation}
where $m$ is the scintillation index at a certain day of observations, $I$ is the current flux density in arbitrary units, $\langle I \rangle $ and is the mean intensity. Performing observations at different elongations, one may derive an estimate of the maximum scintillation index ($m_{max}$) from dependence $m(\epsilon)$ ($\epsilon$ is the source elongation):

%2
\begin{equation}
m_{max}    = x m_0,
\end{equation}
where $x$ is the compactness of the studied source and $m_0$ is the maximum scintillation index corresponding to the compact source component. Scintillation index is, $m_0$ in turn, related to the effective angular size of the source:

%3
\begin{equation}
m_0^2 = [1 + (\theta_1^2 / \theta_{Fr}^2)]^{-1},
\end{equation}
where $\theta_1$ is the effective angular size of the source \citeauthor{Tyulbashev2019} (\citeyear{Tyulbashev2019}) and $\theta_{Fr}$ is the angular size of the Fresnel region that depends on the wavelength (center observational frequency) and the distance to the modulating layer ($\theta_{Fr} =2.06265 \times 10^5 \times (z/k)^{1/2}$), where $\theta_{Fr}$ is expressed in arcseconds, $z$ is the distance to the modulating layer, $k = 2 \pi / \lambda$ is the wavenumber, and $\lambda$ is the wavelength. The effective angular size is related to the actual angular size in the following way:

%4
\begin{equation}
\theta_0 = \beta \theta_1,
\end{equation}
where $\theta_0$ is the angular size of the compact radio source and $\beta = 0.8$ is a coefficient that was determined in \citeauthor{Tyulbashev2019} (\citeyear{Tyulbashev2019}) based on the results of observations of several sources with known angular sizes. The asymmetry coefficient, which is defined as

%5
\begin{equation}
\gamma = \langle (I - \langle I \rangle )^3 \rangle / [\langle (I - \langle I \rangle)^2 \rangle ]^{3/2},
\end{equation}
is proportional to the scintillation index and, according to [6], does not depend on the extended (nonscintillating) component (halo):

%6
\begin{equation}
\gamma = A m_{max}  =A_0 m_0
\end{equation}
$A$ and $A_0$ are the coefficients of proportionality determined in \citeauthor{Tyulbashev2019} (\citeyear{Tyulbashev2019}). Thus, the angular sizes of sources were estimated in \citeauthor{Tyulbashev2019} (\citeyear{Tyulbashev2019}) using the parameters derived from the estimates of the scintillation index at the maximum of $m(\epsilon)$ and parameter $A$, which is calculated based on the asymmetry coefficient:

%7
\begin{equation}
\theta_0 = \beta \theta_{Fr} [(A_0 / A m_{max})^2 - 1 ]^{1/2}  .
\end{equation}

Note that the values of $A$ and $m_{max}$ in (7) are derived from the $\gamma(m)$ and $m(\epsilon)$ dependencies (i.e., from observational data). These dependencies are unique to each source. The value of $\theta_{Fr}$ is calculated under the assumption that the distance to the modulating layer is known. By default, this distance is assumed to be $1 a.u. \times \cos (\epsilon)$. However, this distance is different, e.g., in the case of a coronal mass ejection with the line of sight crossing the ejection. Since the Sun is quiet most of the time, the angular size of the Fresnel region for a given elongation is assumed to be constant. The values of $\beta = 0.8$ and $A_0 = 2$ were determined in \citeauthor{Tyulbashev2019} (\citeyear{Tyulbashev2019}) for a selection of sources with independent estimates of the angular size derived from the results of observations at 103 MHz (\citeauthor{Janardhan1993}, \citeyear{Janardhan1993}). The value of $A_0$ varies from 1.5 to 3 in different models of turbulence of the interplanetary medium (\citeauthor{Mercier1962} (\citeyear{Mercier1962}), \citeauthor{Tatarskii1967} (\citeyear{Tatarskii1967})). $\beta$ is a constant that was determined in the study of sources with known angular sizes \citeauthor{Tyulbashev2019} (\citeyear{Tyulbashev2019}) and that may be used for any sources. Presumably, it is a simple coefficient of proportionality that includes all the possible uncertainties associated with the interplanetary medium. The value of $A_0$ was determined in \citeauthor{Tyulbashev2019} (\citeyear{Tyulbashev2019}) and corresponds to the regions of turbulent solarwind plasma that causes scintillation at 111 MHz (heliocentric distances from 0.4 to 1 a.u.). If the method is used at higher frequencies (specifically, 327 MHz) with the modulating plasma located closer to the Sun, constant $A_0$ should be redefined.

As is known, the spatial distribution of turbulent solar wind at the solar maximum is almost spherically symmetric. At the solar minimum, the plasma density is increased at low heliolatitudes and reduced at high heliolatitudes under the influence of the heliospheric current sheet (\citeauthor{Shishov2016} (\citeyear{Shishov2016}), \citeauthor{Tokumaru2012} (\citeyear{Tokumaru2012}), \citeauthor{Manoharan2012} (\citeyear{Manoharan2012})). The variations are insignificant at low latitudes and maximized at high latitudes. Cyclic evolution of the spatial plasma distribution may lead to changes in $\theta_{Fr}$ and $A_0$ (see (7)), which are calculated under the assumption of spherical symmetry. Since accurate plasma distributions for each year are not known, parameter $\beta$ in (7) is determined in the present study for every year separately to compensate for the possible yearly variations of the indicated parameters.

In our observations, the year 2014 corresponds to the maximum solar activity. In subsequent years, this activity decreased. Quasar 3C 48, which was studied thoroughly in the in the meter wavelength range using both the interplanetary scintillation method and interferometric techniques (\citeauthor{Janardhan1993} (\citeyear{Janardhan1993}), \citeauthor{Glubokova2013} (\citeyear{Glubokova2013}), \citeauthor{Artyukh1999} (\citeyear{Artyukh1999})), was taken as an example to check for the probable influence of the medium on the estimated parameters. Its scintillation index and asymmetry coefficient were estimated for each year of observations. Cases of strong interference, which is not suppressed by the processing software; abrupt changes in the scintillation index associated with coronal mass ejections or corotating flows; and distortion of the source shape, which may result even in its splitting due to ionospheric scintillation, are routinely encountered in data processing. All these factors may lead to the emergence of outliers on the $m(\epsilon)$  dependence. These outliers were removed by the processing software. Alongside the indicated factors, a series of short-duration interference signals, which exert almost no influence on the determined flux-density fluctuations, may produce a significant contribution to the asymmetry coefficient. Therefore, both outlier scintillation index values and large deviations of the asymmetry coefficient were excluded from the $\gamma(m)$ dependence. Figure 2 shows pairs of dependencies $m(\epsilon)$ and $\gamma(m)$ for 3C 48 for 2014–2018.

Using the least-squares method, dependencies $m(\sin(\epsilon))$ and $\gamma(m)$ were approximated by logarithmic and common straight lines, respectively. The obtained $m_{max}$ and $A$ estimates provide an opportunity to estimate the compactness and angular size of 3C 48 in accordance with (7) under the assumption that $\beta = 0.8$. These values are listed in Table~\ref{tab:tab1}.

\begin{figure*}
\begin{center}
	\includegraphics[width=0.9\textwidth]{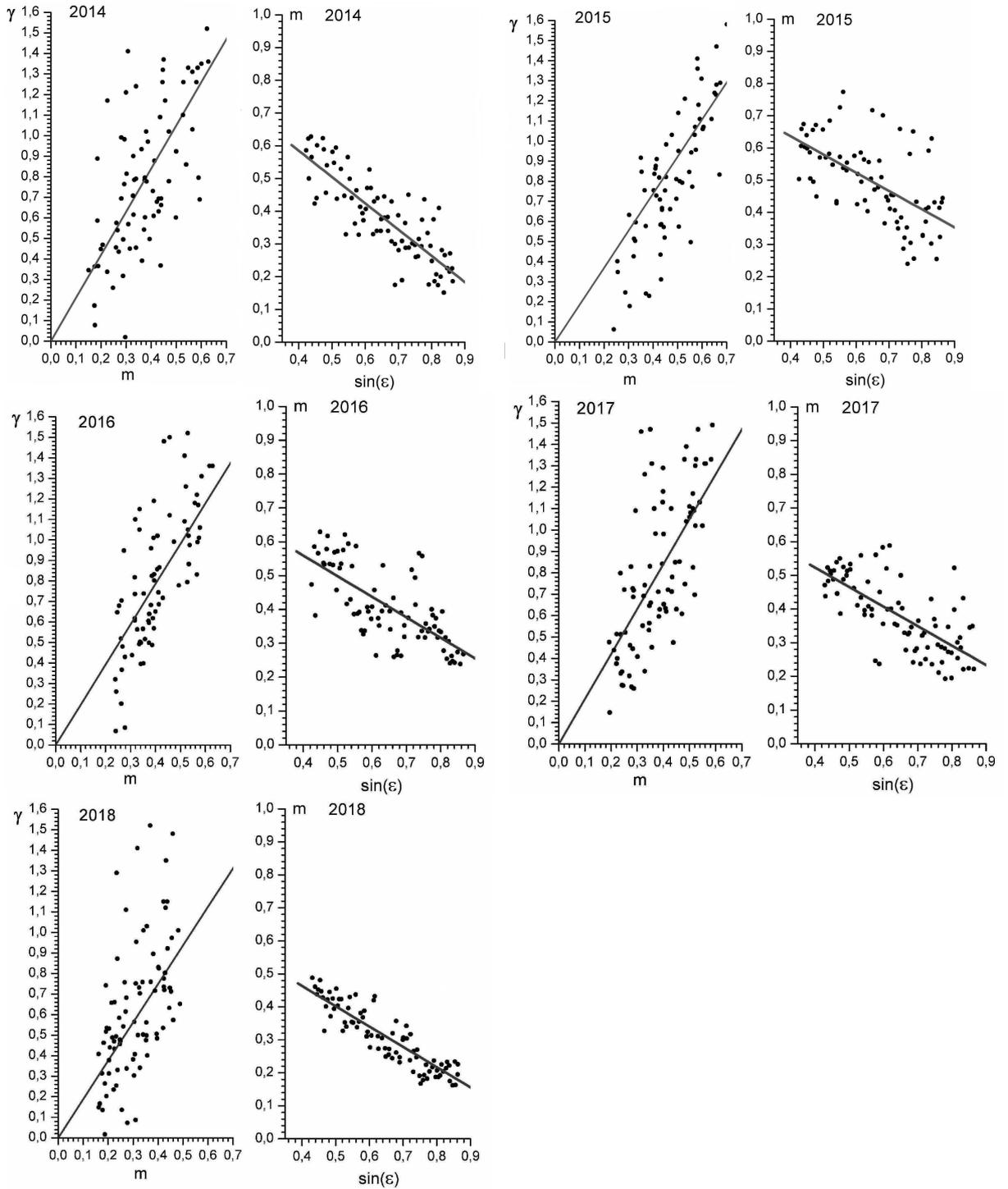}
    \caption{ Pairs of dependencies of observed scintillation index $m$ on elongation sine and asymmetry coefficient $\gamma$ on the scintillation index. The year of observations is indicated at the top for each pair. The straight line is the result of least-squares approximation. It can be seen that the maximum scintillation index of 3C 48 at elongations most suitable for observation of scintillations (near $\epsilon=25^{\circ}$) varies from one year to the other and falls within the 0.5–0.65 interval. It is also evident that the slope of the approximating line, which defines coefficient , also varies from one year to the other and falls within the interval from 0.93 to 1.05.}
    \label{fig:fig2_TChSK}
\end{center}
\end{figure*}

\begin{table}
	\centering
	\caption{Angular size, maximum scintillation index, and compactness of quasar 3C 48 in 2014–2018}
	\label{tab:tab1}
	\begin{tabular}{cccccc}
		\hline
Year & 2014 & 2015 & 2016 & 2017 & 2018\\
		\hline
$\theta_0$, arcsec & 0.29  & 0.41     & 0.34     & 0.42     & 0.53\\ 
$m_{max}$ & 0.64  & 0.66     & 0.63     & 0.57     & 0.52\\
$x$  & 0.91  & 1.14     & 0.96     & 0.99     & 1.06\\
    	\hline
	\end{tabular}
	\label{tab:tab1}
\end{table}

It can be seen that the compactness parameter of 3C 48 is close to unity; i.e., the accuracy of compactness determination was no worse than $\pm10\%$, and the flux-density contribution of the extended (nonscintillating) component of 3C 48 is small or nonexistent. The angular size of 3C 48 increases starting from 2014, while the maximum scintillation index decreases with time. Note once again that the year 2014 corresponds to the maximum solar activity, while the activity in 2018 was near its minimum.

\begin{figure}
	\includegraphics[width=0.9\columnwidth]{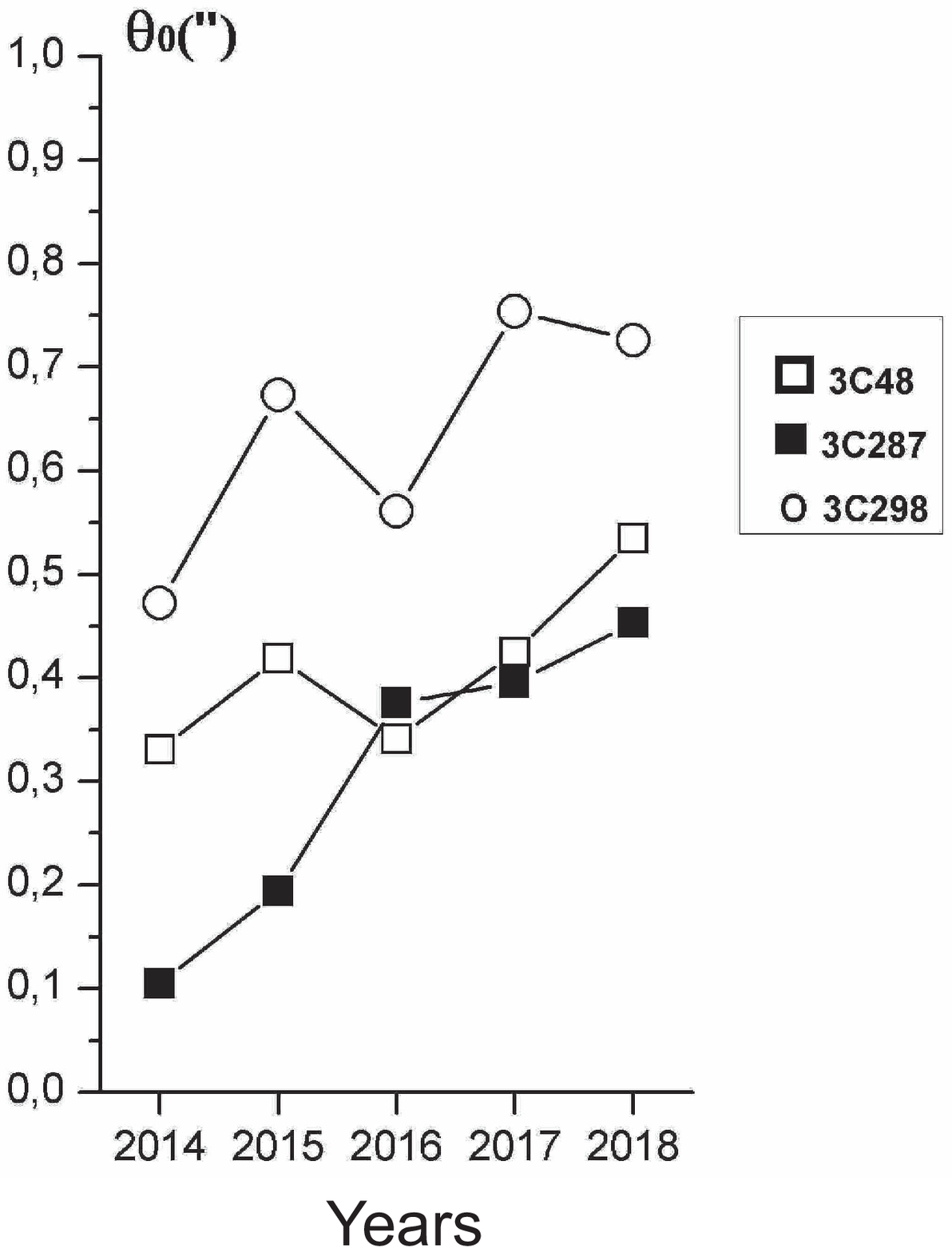}
	\caption{Temporal variations of angular sizes $\theta_0$ of 3C 48, 3C 287, and 3C 298 induced by the reduction in the level of small-scale fluctuations of the plasma concentration averaged over the probed solar-wind region.}
    \label{fig:fig3_TChSK}
\end{figure}

Estimates of the angular sizes were also obtained for two radio sources that were used in \citeauthor{Tyulbashev2019} (\citeyear{Tyulbashev2019}) as the basis for estimating the angular sizes of the entire sample of strong scintillating sources (see Fig.~\ref{fig:fig3_TChSK}, which shows the size estimates for different years of observation determined under the assumption that $\beta = 0.8$ ). Independent size estimates for this pair of sources were obtained based on the results of observations at 103 MHz in (\citeauthor{Janardhan1993}, \citeyear{Janardhan1993}).

On cosmological scales, an angular size of a fraction of a second corresponds to a linear size varying from several hundred parsec to several kiloparsec. It is evident that the angular size cannot increase by a factor of 2 due to changes in some structures in the studied sources occurring within several years. Therefore, it is rational to assume that the growth of the estimated angular size of sources with time is attributable to certain changes in the interplanetary plasma.

The components of expression (7), which is used to calculate the angular size, were discussed in some detail directly below this formula. Since the angular size of sources cannot increase physically, it is logical to assume that these changes are related to the 11-year solar cycle, and it seems most probable that the size estimate of the Fresnel region used changes from one year to another. Thus, the estimate of angular size $\theta_{Fr}$ of the Fresnel region in the in the meter wavelength range in different phases of the solar cycle is, in general, not entirely correct. In the case of a spherically symmetric solar-wind distribution, which is found at the solar maximum, the modulation of radio waves is maximized near the line-of-sight aiming point. The contribution of low-latitude regions, where the plasma concentration is elevated relative to that in high-latitude regions, increases as the activity minimum draws nearer. As a result, the modulating region shifts away from the aiming point toward the observed for sources with the line of sight to them going through middle and high heliolatitudes.

The Fresnel size increases with distance to the modulating layer. This implies that when the distance to the modulating layer decreases, scintillations are suppressed partially. In estimates obtained using the method from \citeauthor{Tyulbashev2019} (\citeyear{Tyulbashev2019}), this will be manifested formally as an increase in the angular sizes of sources. When applying formula (7) in practice, it is more convenient to assume that coefficient $\beta$ in (4) and (7), which needs to be determined for each year individually, changes instead of the Fresnel size. This approach allows one to eliminate all uncertainties in coefficient $\beta$. The studied radio sources are observed at optimum elongations at different times in the course of a year, and if the global solar wind structure changes significantly from one month to another, coefficient $\beta$ varies together with it. These intrayear variations are expected to be insignificant, since they do not exceed the interyear ones.

All the available estimates of the angular sizes of sources corresponding to different years were used to estimate correctly the relative variation of from one year to another. With this end in view, the mean angular size was calculated for each source, and the size estimates for 2014–2018 were divided by this angular size. This allows one to renormalize angular sizes so as to make the mean angular size of each source equal to 1$^{\prime\prime}$. The median angular size was then estimated for each year individually, and root-mean-square deviations from this size were estimated. The same was done for the maximum scintillation indices. Dependencies $m(\epsilon)$ and $\gamma(m)$ were plotted for elongations ranging from $25^{\circ}$ to $60^{\circ}$. The mean number of points was 69 and 63 in dependencies $m(\epsilon) = 69$ and $\gamma(m) = 63$, respectively. The median number of points in these dependences was 68 and 62. The corresponding root-mean square deviations are $\sigma_m = 0.3$ and $\sigma_{\gamma} = 0.3$.

\begin{figure*}
\begin{center}
	\includegraphics[width=0.9\textwidth]{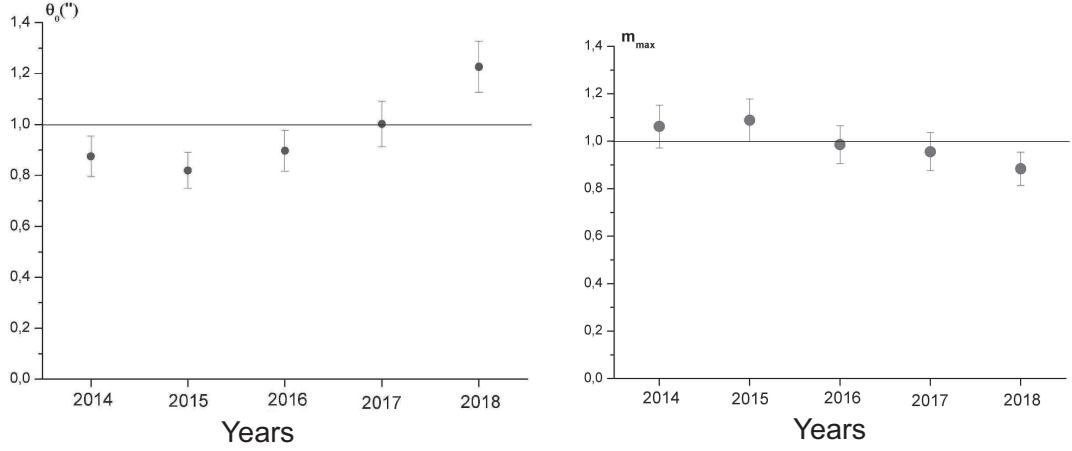}
    \caption{ Dependencies of the angular size (left) and the maximum scintillation index (right) of a source. Vertical bars denote the root-mean-square deviations for each year.}
    \label{fig:fig4_TChSK}
\end{center}
\end{figure*}

Fig.~\ref{fig:fig4_TChSK} and Table~\ref{tab:tab2} present the time dependencies of relative variations of angular sizes and scintillation indices.

\begin{table*}
	\centering
	\caption{Relative variations and errors of normalized mean angular sizes and maximum scintillation indices in 2014–2018}
	\label{tab:tab2}
	\begin{tabular}{cccccc}
		\hline
Year & 2014 & 2015 & 2016 & 2017 & 2018\\
$\theta_0$, arcsec & 0.88$\pm$0.08 & 0.82$\pm$0.07 & 0.90$\pm$0.08 & 1.00$\pm$0.09 & 1.23$\pm$0.10\\ 
$m_{max}$ & 1.06$\pm$0.09 & 1.09$\pm$0.09 & 0.97$\pm$0.08 & 0.96$\pm$0.08 & 0.88$\pm$0.07\\
       \hline
    \end{tabular}
	\label{tab:tab2}
\end{table*}

It follows from Fig.~\ref{fig:fig4_TChSK} that the value of $\beta$  changed by a factor of $\sim$1.4 within the interval from 2014 to 2018.

Having determined the nature of relative variation of coefficient $\beta$ with time and assuming that the angular size of 3C 48 should be 0.33$^{\prime\prime}$ (\citeauthor{Janardhan1993} (\citeyear{Janardhan1993}), \citeauthor{Glubokova2013} (\citeyear{Glubokova2013}), \citeauthor{Artyukh1999} (\citeyear{Artyukh1999})), we redefined the values of $\beta$ after averaging the estimates of the angular source size over 5 years. These resulting $\beta$ values, which are listed in Table~\ref{tab:tab3}, were used subsequently to estimate the angular sizes of compact sources.

\begin{table}
	\centering
	\caption{Estimates of coefficient in 2014–2018}
	\label{tab:tab3}
	\begin{tabular}{cccccc}
		\hline
Year & 2014 & 2015 & 2016 & 2017 & 2018\\
		\hline
$\beta$ & 0.72  & 0.77     & 0.70     & 0.63     & 0.51\\ 
    	\hline
	\end{tabular}
	\label{tab:tab3}
\end{table}

\section{DISCUSSION}

The Pushchino catalogue of discrete radio sources (\citeauthor{Dagkesamanskii2000}, \citeyear{Dagkesamanskii2000}) was compiled based on the results of a BSA LPI survey performed at 102.5 MHz in 1991–1993. A number of sources with known flux densities were used for calibration in this survey. Catalogue (\citeauthor{Slee1995}, \citeyear{Slee1995}) based on the observations made at 80 and 160 MHz was used to check for the presence of systematic errors in \citeauthor{Dagkesamanskii2000} (\citeyear{Dagkesamanskii2000}). With this end in view, the flux density of sources at 80 and 160 MHz was interpolated linearly to 102.5MHz, and the dependence of the observed flux density at 102.5 MHz on the interpolated flux density at the same frequency was then plotted to reveal significant systematic errors. Although the flux densities measured for certain radio sources by the BSA LPI telescope differ from the interpolated values by a factor of 1.5–2, no systematic deviations were found.

Monitoring observations were conducted using the same BSA LPI antenna, but with the center frequency changed after reconstruction. Since the frequencies of 102.5 and 110.3 MHz differ by less than $10\%$, corrections for the flux densities from the catalogue for 102.5MHz should be small. According to the figures from the paper detailing the results of a survey at 151 MHz (\citeauthor{McGilchrist1990}, \citeyear{McGilchrist1990}), approximately $90\%$ of all sources observed at 151 MHz have spectral indices falling within the range $0.6 < \alpha <1.2 (S \sim \nu^{-\alpha})$. Therefore, the extreme cases for these $90\%$ of sources are $\alpha = 0.6$ and $\alpha = 1.2$. Assuming that all the studied sources have $\alpha = 0.9$, we recalculated their flux densities to 110.3MHz and used the obtained flux densities for self-calibration. The errors of $m_{max}$  determination of the flux density at 110.3 MHz associated with the use of $\alpha = 0.9$ for all sources instead of individual spectral indices (derived from the frequency dependence of the flux  density) do not exceed $\pm4\%$ for $90\%$ of sources. Since the errors of estimates come up approximately to $\pm10\%$ from year to year, errors introduced into the estimate of the integrated flux density at 111 MHz may be neglected.

The estimation of the angular size and compactness of sources was discussed in the above paragraphs. An estimate of the maximum scintillation index ($m_{max}$) was derived as an intermediate result for each source. Since sources with flux densities above 5 Jy are studied, individual signal calibration for each source (i.e., self-calibration) may be used to estimate fluctuations of the flux density expressed in janskys. Specifically, self-calibration is convenient in that it allows one to ignore the possible shift of coordinates of the source due to ionospheric scintillations and corrections for the meridian attitude of the source and to skip the calculation of the contribution to the observed flux density due to the declination coordinate of the source falling between the neighboring BSA LPI beams. The primary errors arising in the use of self-calibration will be related to systematic errors of the catalogue used as the basis.

Estimates of the scintillation index ($m_{max}, m_0$), angular size ($\theta_0$), compactness ($x$), flux-density fluctuations ($\Delta S$), and flux density of the compact component ($S_c$) obtained independently for each year of monitoring observations were averaged. The results for all sources are given in the Appendix. The angular sizes of sources exceeding 1 arcsec were excluded from Table A1 (see Appendix), since it was demonstrated in \citeauthor{Tyulbashev2019} (\citeyear{Tyulbashev2019}) that such sizes are determined incorrectly. The estimates of $m_{max}$ and $\Delta S$ were obtained for all the remaining sources.

\begin{figure*}
\begin{center}
	\includegraphics[width=0.9\textwidth]{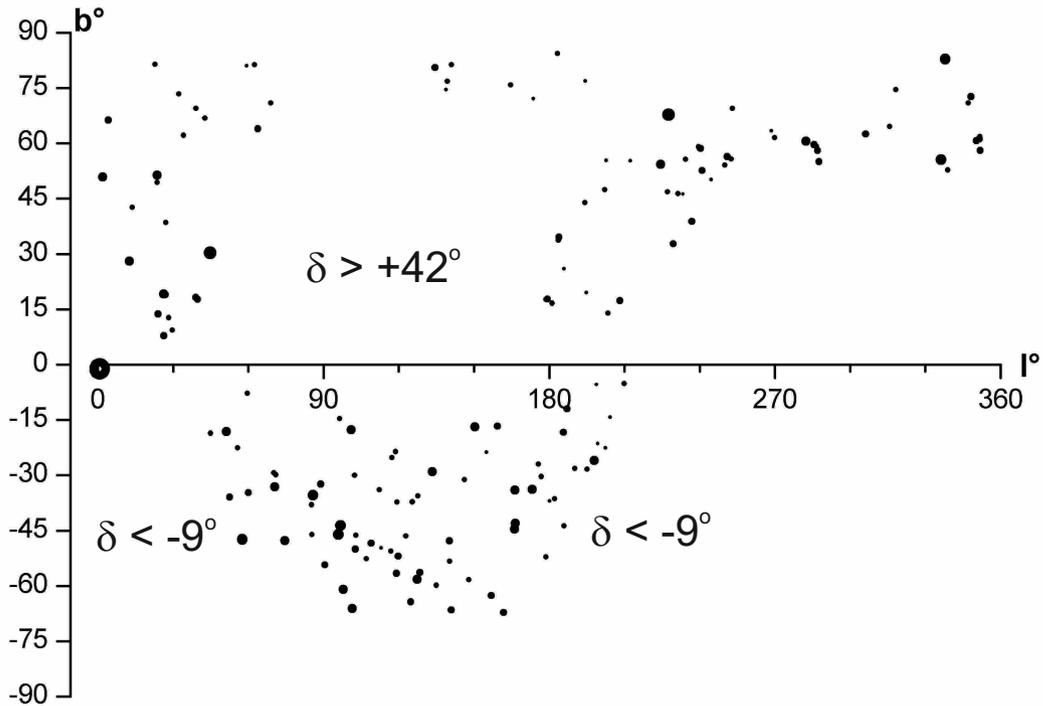}
    \caption{ All sources with estimates of their angular sizes available. The circle size is proportional to the angular size of a source. The large circle near $b = 0^{\circ}$, $l = 0^{\circ}$ denotes the Galaxy center. Large declination regions are left blank, since they are beyond the reach of monitoring observations. A blank stripe running along the Galaxy plane ($b = 0^{\circ}$) illustrates the lack of sources with angular sizes less then 1$^{\prime\prime}$ found within $\pm5^{\circ}$ of the Galaxy plane.}
    \label{fig:fig5_TChSK}
\end{center}
\end{figure*}

Fig.~\ref{fig:fig5_TChSK} shows 178 sources with their angular sizes identified. It can be seen that no sources smaller than 1$^{\prime\prime}$ are found in the Galactic plane. This is hardly surprising, since the signal is scattered in the dense interstellar medium and the observed angular sizes of compact radio sources increase as one approaches the Galactic plane.

\begin{figure*}
\begin{center}
	\includegraphics[width=0.9\textwidth]{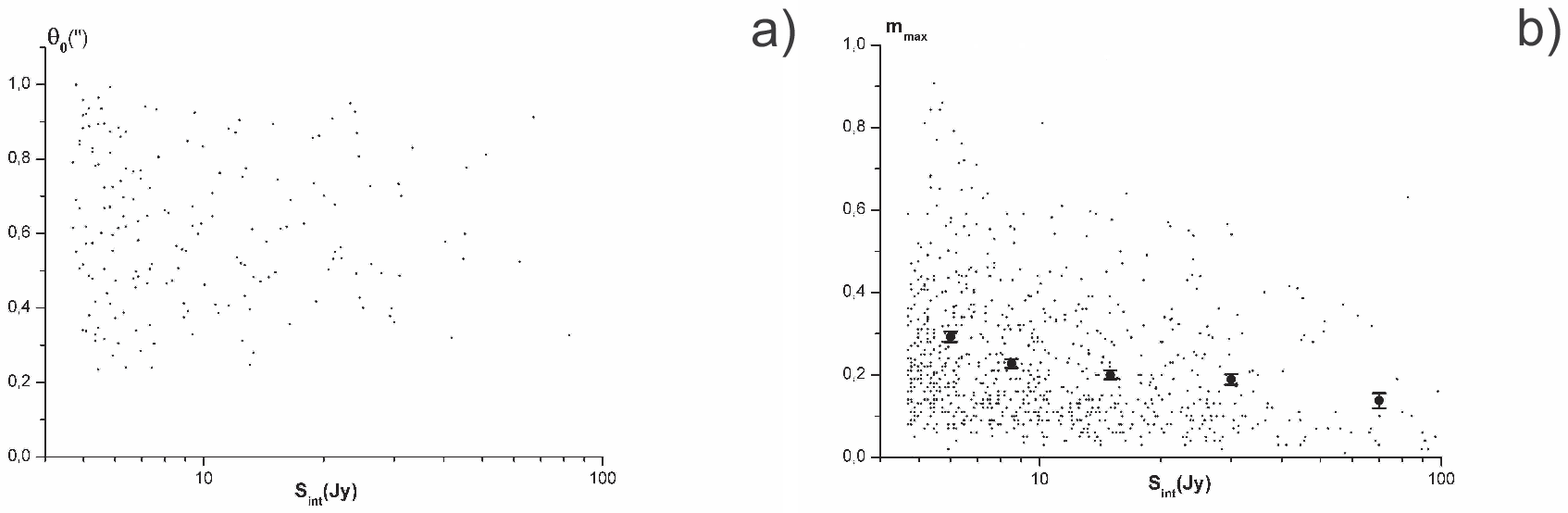}
    \caption{ Dependences of (a) observed angular size $\theta_0$ and (b) observed maximum scintillation index $m$ at $\epsilon = 25^{\circ}$ of sources on their integrated flux density (in logarithmic scale). Large circles and bars denote the mean values and their root-meansquare deviations $m_{max}$ within the following flux-density intervals: $4.5 < S_{int} \le 7$; $7 < S_{int} \le 10$; $10 < S_{int} \le 20$; $20 < S_{int} \le 40$; $S_{int} > 40$ Jy.}
    \label{fig:fig6_TChSK}
\end{center}
\end{figure*}

As was already noted, 5 Jy is the boundary flux density that ensures completeness of the survey. One may examine the relations between the determined parameters of sources with compact (scintillating) components and the integrated flux density of the complete sample of sources. If these compact components are smaller than 1$^{\prime\prime}$, the possible relation between the angular size of a compact radio source and other determined parameters may also be examined. We checked whether the compactness and the angular size, the compactness and the integrated flux density, the flux density of the compact component and the integrated flux density, and the angular size and the integrated flux density of sources are interrelated. No marked dependencies were found. Figure 6a illustrates this point: it can be seen that the angular size of sources and their integrated flux density are not related in any significant way. Fig.~\ref{fig:fig6_TChSK}b illustrates the relation between the maximum scintillation index and the integrated flux density.

From Fig.~\ref{fig:fig6_TChSK}a it is seen that there is no explicit relationship between the observed flux density and the angular dimensions of the source. At the same time, in Fig.~\ref{fig:fig6_TChSK}b there is a dependence, a clear $m_{max} / S_{int}$ dependence is seen: weaker sources tend to have higher scintillation indices.

Let us also mention a number of restrictions on the applicability of the developed method for estimating the angular sizes of sources. First, it is applicable to sources with angular sizes less then 1$^{\prime\prime}$. Scattering sets boundaries for its use at galactic latitudes of $\pm5^{\circ}$ (see Fig.~\ref{fig:fig5_TChSK}). Second, the obtained estimates of angular sizes were found to be related to the solar activity in the 11-year cycle. Separate estimates of coefficient $\beta$ will be needed in observations at different frequencies. Third, estimates of angular sizes may be derived from the scintillation power spectrum and elongation dependencies of the scintillation index in the region of weak and strong scintillation. Since the interplanetary medium changes from one year to the other, its influence will be manifested in all these methods and needs to be taken into account.

\section{CONCLUSIONS}

Estimates of the angular size of compact radio sources obtained using the asymmetry coefficient vary from one phase of the solar activity cycle to the other. At the solar maximum, the spherical component prevails and correction coefficient $\beta$ (see formula (4)), which relates the true angular size to the size determined using the asymmetry coefficient, is close to 0.75. If the angular size is estimated at the solar minimum, the equatorial component is significant, and coefficient $\beta$ is close to 0.5.

This relation between the size estimate of a scintillating radio source and the phase of the solar activity cycle may depend on the frequency of observations. Variations in the decimeter range are expected to be close to or stronger than those in the in the meter wavelength range, since high heliolatitudes produce a significant contribution to the observational data. At the same time, variations in the decameter range should be relatively weak, since the primary contribution to scintillations is produced by low-latitude solar-wind regions close to the orbit of the Earth, which feature weak cyclic variations.

Note that other methods for size estimation relying on the assumption of spherical symmetry of the solarwind distribution should also feature a relation between the angular size estimate and the solar activity phase, since the symmetry condition is present implicitly in the formulas used to estimate the angular size based both on the elongation dependence of the scintillation index and on the power spectra. An asymmetric distribution of interplanetary plasma will result in overestimation of the angular size in the years with low solar activity levels.

The obtained catalogue of compact radio sources (see Appendix) may be used as a basis to study the solar activity using observational data on scintillations of radio sources. At the same time, this catalogue in itself is valuable in studies of extragalactic radio sources, since estimates of the flux densities and angular sizes for complete samples of compact radio sources are scarce.

%\section{APPENDIX \\ CATALOGUE OF COMPACT RADIO SOURCES}

%The designation of a source given in the Pushchino catalogue of discrete sources is indicated in the first column. The following parameters averaged over a 5-year interval are listed in columns 2–6: maximum scintillation index ($m_{max}$), angular size of a source ($\theta_0$), its compactness (fraction of energy in the compact component, $x$), observed flux-density fluctuations at the scintillation maximum ($\Delta S_{max}$), and flux density of the compact component ($S_c$) determined in accordance with (2). The integrated flux density ($S_{int}$) recalculated from 102.5 to 110.3 MHz under the assumption that all the studied sources have a spectral index of 0.9 is given in the seventh column. A dash indicates that the angular size of a certain source could not be determined; other parameters requiring an angular size value to be determined are also not indicated for such sources.
%\newpage
%\input{TChSKtable.tex}

\section{FUNDING}

The work was supported by the Program of the Presidium of the Russian Academy of Sciences KP19-270 "Questions of the origin and evolution of the Universe using methods of ground-based observations and space research"

% Don't change these lines
\bsp	% typesetting comment

\section{APPENDIX \\ CATALOGUE OF COMPACT RADIO SOURCES}

The designation of a source given in the Pushchino catalogue of discrete sources is indicated in the first column. The following parameters averaged over a 5-year interval are listed in columns 2–6: maximum scintillation index ($m_{max}$), angular size of a source ($\theta_0$), its compactness (fraction of energy in the compact component, $x$), observed flux-density fluctuations at the scintillation maximum ($\Delta S_{max}$), and flux density of the compact component ($S_c$) determined in accordance with (2). The integrated flux density ($S_{int}$) recalculated from 102.5 to 110.3 MHz under the assumption that all the studied sources have a spectral index of 0.9 is given in the seventh column. A dash indicates that the angular size of a certain source could not be determined; other parameters requiring an angular size value to be determined are also not indicated for such sources.
\newpage
%\input{TChSKtable.tex}

%\tablecaption{CATALOGUE OF COMPACT RADIO SOURCES }
\tablehead{
\hline
Name	&	$m_{m}$	&	$\theta_0  ('')$	&	$x$	&	$\Delta S_{m}$ (Jy)	&	$S_c$~ (Jy)	&	$S_{int}$ (Jy)	\\
\hline
}
\tabletail{\hline}
\begin{supertabular}{|c|c|c|c|c|c|c|}
0002+126	&	0.08	&	-	&	-	&	2.4	&	-	&	30.5	\\
0003+196	&	0.69	&	-	&	-	&	4	&	-	&	5.8	\\
0004-004	&	0.44	&	0.4	&	0.88	&	11	&	22.2	&	25.1	\\
0004+218	&	0.15	&	-	&	-	&	0.9	&	-	&	6.3	\\
0004+382	&	0.2	&	-	&	-	&	2.8	&	-	&	13.8	\\
0007+124	&	0.43	&	-	&	-	&	7.8	&	-	&	18.1	\\
0010+353	&	0.11	&	-	&	-	&	0.6	&	-	&	5.4	\\
0010+348	&	0.19	&	-	&	-	&	0.9	&	-	&	4.9	\\
0010+344	&	0.19	&	-	&	-	&	1.6	&	-	&	8.5	\\
0010+266	&	0.27	&	-	&	-	&	1.3	&	-	&	5	\\
0011+281	&	0.32	&	0.44	&	0.58	&	1.8	&	3.3	&	5.7	\\
0011+206	&	0.12	&	-	&	-	&	2.2	&	-	&	11.9	\\
0012+322	&	0.1	&	-	&	-	&	2.7	&	-	&	26.7	\\
0012+090	&	0.36	&	0.61	&	1	&	2.2	&	6.1	&	6.1	\\
0013+135	&	0.58	&	0.41	&	1	&	6.2	&	10.7	&	10.7	\\
0015+214	&	0.06	&	-	&	-	&	0.4	&	-	&	6.9	\\
0017-047	&	0.5	&	0.86	&	1	&	3.1	&	6.2	&	6.2	\\
0017+154	&	0.12	&	-	&	-	&	2.8	&	-	&	23.4	\\
0018+242	&	0.22	&	-	&	-	&	1.2	&	-	&	5.3	\\
0019-086	&	0.36	&	-	&	-	&	2.4	&	-	&	6.7	\\
0025+126	&	0.19	&	0.61	&	0.53	&	3	&	8.3	&	15.6	\\
0025+373	&	0.24	&	0.67	&	0.7	&	1.2	&	3.4	&	4.9	\\
0026+258	&	0.2	&	-	&	-	&	1.1	&	-	&	5.4	\\
0029+216	&	0.27	&	-	&	-	&	2.5	&	-	&	9.4	\\
0031+063	&	0.15	&	-	&	-	&	2.1	&	-	&	14.3	\\
0031+391	&	0.25	&	0.81	&	0.77	&	6.1	&	18.8	&	24.5	\\
0034+386	&	0.11	&	-	&	-	&	1.2	&	-	&	11.1	\\
0035+121	&	0.27	&	0.72	&	0.88	&	1.6	&	5.2	&	5.9	\\
0035+339	&	0.09	&	-	&	-	&	0.9	&	-	&	10.1	\\
0036+043	&	0.28	&	-	&	-	&	5.7	&	-	&	20.2	\\
0038+329	&	0.14	&	-	&	-	&	2.7	&	-	&	19	\\
0038+088	&	0.12	&	-	&	-	&	1.7	&	-	&	14	\\
0038+255	&	0.25	&	0.61	&	0.67	&	3.4	&	8.9	&	13.2	\\
0039+321	&	0.33	&	-	&	-	&	3.1	&	-	&	9.5	\\
0040+373	&	0.17	&	-	&	-	&	1.9	&	-	&	11.4	\\
0040+281	&	0.09	&	-	&	-	&	0.9	&	-	&	9.5	\\
0042+062	&	0.32	&	0.49	&	0.74	&	1.7	&	3.8	&	5.2	\\
0043+108	&	0.43	&	0.48	&	0.94	&	2.3	&	5	&	5.3	\\
0044+302	&	0.17	&	-	&	-	&	1	&	-	&	5.8	\\
0047+324	&	0.1	&	-	&	-	&	0.5	&	-	&	4.9	\\
0049+118	&	0.23	&	-	&	-	&	1.9	&	-	&	8.2	\\
0050+381	&	0.31	&	-	&	-	&	3	&	-	&	9.7	\\
0051+129	&	0.14	&	-	&	-	&	1.4	&	-	&	10.2	\\
0051+163	&	0.25	&	0.66	&	0.76	&	2	&	6	&	8	\\
0051+169	&	0.13	&	-	&	-	&	0.9	&	-	&	6.9	\\
0054-015	&	0.39	&	0.6	&	0.87	&	17.4	&	39.3	&	45.2	\\
0054+090	&	0.19	&	-	&	-	&	1	&	-	&	5.1	\\
0056+315	&	0.81	&	-	&	-	&	4.2	&	-	&	5.2	\\
0058+106	&	0.16	&	-	&	-	&	1.5	&	-	&	9.5	\\
0059+145	&	0.1	&	-	&	-	&	2.3	&	-	&	23.2	\\
0100+256	&	0.32	&	0.52	&	0.69	&	3.9	&	8.6	&	12.4	\\
0100+045	&	0.65	&	0.34	&	0.97	&	4.4	&	6.6	&	6.8	\\
0104+322	&	0.16	&	-	&	-	&	5.4	&	-	&	34	\\
0105+267	&	0.07	&	-	&	-	&	0.4	&	-	&	5.9	\\
0106+130	&	0.06	&	-	&	-	&	4	&	-	&	66.3	\\
0106+144	&	0.08	&	-	&	-	&	0.6	&	-	&	7.9	\\
0108+134	&	0.08	&	-	&	-	&	0.4	&	-	&	5.5	\\
0108+352	&	0.13	&	-	&	-	&	0.6	&	-	&	4.7	\\
0108+272	&	0.24	&	0.62	&	0.53	&	1.6	&	3.4	&	6.4	\\
0110+296	&	0.2	&	-	&	-	&	0.9	&	-	&	4.7	\\
0111+339	&	0.18	&	-	&	-	&	0.9	&	-	&	5.1	\\
0115+025	&	0.2	&	0.83	&	0.7	&	1.9	&	7	&	10	\\
0119+379	&	0.06	&	-	&	-	&	0.8	&	-	&	13.2	\\
0119-047	&	0.4	&	0.82	&	1	&	2.3	&	5.8	&	5.8	\\
0124+231	&	0.28	&	-	&	-	&	1.7	&	-	&	5.9	\\
0124+347	&	0.23	&	-	&	-	&	1.8	&	-	&	7.9	\\
0124+326	&	0.09	&	-	&	-	&	1.3	&	-	&	14.6	\\
0125+288	&	0.1	&	-	&	-	&	1.9	&	-	&	19.5	\\
0127+234	&	0.32	&	-	&	-	&	5.8	&	-	&	18.1	\\
0128+251	&	0.14	&	-	&	-	&	2.5	&	-	&	17.6	\\
0128+063	&	0.13	&	-	&	-	&	2	&	-	&	15.7	\\
0133+206	&	0.07	&	-	&	-	&	3.7	&	-	&	52.3	\\
0134+310	&	0.26	&	-	&	-	&	1.3	&	-	&	5.2	\\
0134+329	&	0.63	&	0.33	&	0.98	&	52	&	80.6	&	82.5	\\
0135+227	&	0.38	&	-	&	-	&	4.1	&	-	&	10.9	\\
0138+135	&	0.58	&	0.5	&	1	&	8.7	&	15.1	&	15.1	\\
0139+295	&	0.08	&	-	&	-	&	0.4	&	-	&	5.3	\\
0140+388	&	0.32	&	-	&	-	&	2	&	-	&	6.4	\\
0141+341	&	0.21	&	-	&	-	&	1.7	&	-	&	7.9	\\
0142+020	&	0.22	&	0.85	&	1	&	1.1	&	4.9	&	4.9	\\
0146+064	&	0.09	&	-	&	-	&	1.1	&	-	&	12	\\
0147+268	&	0.17	&	-	&	-	&	0.9	&	-	&	5.4	\\
0151-038	&	0.53	&	0.8	&	1	&	4.1	&	7.7	&	7.7	\\
0151+407	&	0.36	&	-	&	-	&	4.6	&	-	&	12.8	\\
0152+039	&	0.1	&	-	&	-	&	3.2	&	-	&	32	\\
0153+286	&	0.13	&	-	&	-	&	3.2	&	-	&	24.3	\\
0154+209	&	0.1	&	-	&	-	&	1.1	&	-	&	11.4	\\
0154+224	&	0.15	&	-	&	-	&	2	&	-	&	13.3	\\
0157+011	&	0.34	&	-	&	-	&	2.2	&	-	&	6.4	\\
0203+292	&	0.11	&	-	&	-	&	1.4	&	-	&	13.2	\\
0204+295	&	0.11	&	-	&	-	&	1	&	-	&	9.4	\\
0208+211	&	0.09	&	-	&	-	&	1.1	&	-	&	12	\\
0210+049	&	0.2	&	-	&	-	&	4.9	&	-	&	24.7	\\
0211+119	&	0.17	&	-	&	-	&	1.7	&	-	&	9.8	\\
0212+349	&	0.2	&	-	&	-	&	1.9	&	-	&	9.5	\\
0215+108	&	0.35	&	-	&	-	&	3.6	&	-	&	10.3	\\
0216+022	&	0.43	&	-	&	-	&	10	&	-	&	23.3	\\
0218+111	&	0.38	&	-	&	-	&	2.9	&	-	&	7.6	\\
0219+398	&	0.07	&	-	&	-	&	3.4	&	-	&	48.9	\\
0221+276	&	0.43	&	0.43	&	0.82	&	5.5	&	10.3	&	12.7	\\
0222+267	&	0.15	&	-	&	-	&	1.1	&	-	&	7.5	\\
0225+370	&	0.24	&	-	&	-	&	2.8	&	-	&	11.8	\\
0226+344	&	0.11	&	-	&	-	&	0.6	&	-	&	5.7	\\
0226+294	&	0.14	&	-	&	-	&	1.9	&	-	&	13.5	\\
0227+307	&	0.36	&	-	&	-	&	1.7	&	-	&	4.8	\\
0228+344	&	0.1	&	-	&	-	&	2.4	&	-	&	23.9	\\
0229+375	&	0.17	&	-	&	-	&	1.3	&	-	&	7.4	\\
0231+132	&	0.19	&	-	&	-	&	1.2	&	-	&	6.3	\\
0231+314	&	0.13	&	-	&	-	&	3.7	&	-	&	28.1	\\
0232+229	&	0.11	&	-	&	-	&	1	&	-	&	8.7	\\
0235+272	&	0.32	&	-	&	-	&	2.9	&	-	&	8.9	\\
0235+099	&	0.1	&	-	&	-	&	1.6	&	-	&	16.1	\\
0238+100	&	0.12	&	-	&	-	&	0.8	&	-	&	6.3	\\
0238+089	&	0.21	&	-	&	-	&	1.1	&	-	&	5.1	\\
0242+365	&	0.09	&	-	&	-	&	1.1	&	-	&	12.6	\\
0242+350	&	0.14	&	-	&	-	&	0.7	&	-	&	4.8	\\
0244+378	&	0.26	&	-	&	-	&	1.4	&	-	&	5.4	\\
0244+282	&	0.14	&	-	&	-	&	2	&	-	&	14.1	\\
0246+334	&	0.16	&	-	&	-	&	1.7	&	-	&	10.6	\\
0249+081	&	0.91	&	0.35	&	1	&	4.9	&	5.4	&	5.4	\\
0250+385	&	0.08	&	-	&	-	&	1	&	-	&	12.1	\\
0250-011	&	0.31	&	-	&	-	&	1.7	&	-	&	5.4	\\
0253-030	&	0.4	&	0.87	&	1	&	2.6	&	6.4	&	6.4	\\
0254+093	&	0.68	&	0.33	&	1	&	3.7	&	5.4	&	5.4	\\
0258+349	&	0.23	&	0.75	&	1	&	2.8	&	12.5	&	12.5	\\
0300+375	&	0.15	&	-	&	-	&	2.3	&	-	&	15.1	\\
0300+276	&	0.13	&	-	&	-	&	0.9	&	-	&	7.1	\\
0301+335	&	0.16	&	-	&	-	&	0.8	&	-	&	5.2	\\
0303+349	&	0.21	&	-	&	-	&	1	&	-	&	4.8	\\
0307+252	&	0.4	&	-	&	-	&	2.7	&	-	&	6.9	\\
0308+385	&	0.55	&	0.5	&	1	&	4.8	&	8.6	&	8.6	\\
0308+305	&	0.54	&	0.39	&	0.91	&	5.9	&	10	&	10.9	\\
0309+153	&	0.19	&	-	&	-	&	2	&	-	&	10.3	\\
0309+262	&	0.09	&	-	&	-	&	0.9	&	-	&	10.4	\\
0309+048	&	0.5	&	-	&	-	&	7.9	&	-	&	15.9	\\
0310+221	&	0.09	&	-	&	-	&	0.7	&	-	&	7.5	\\
0310+363	&	0.26	&	-	&	-	&	1.2	&	-	&	4.7	\\
0312+318	&	0.09	&	-	&	-	&	0.9	&	-	&	10.4	\\
0317+163	&	0.54	&	0.25	&	0.79	&	7	&	10.3	&	13.1	\\
0317+306	&	0.3	&	-	&	-	&	1.9	&	-	&	6.2	\\
0320+119	&	0.41	&	-	&	-	&	2.7	&	-	&	6.6	\\
0322+358	&	0.13	&	-	&	-	&	1	&	-	&	7.9	\\
0322-035	&	0.36	&	-	&	-	&	4.7	&	-	&	13.1	\\
0325+290	&	0.1	&	-	&	-	&	0.5	&	-	&	5	\\
0328+248	&	0.3	&	-	&	-	&	1.8	&	-	&	6.1	\\
0328+294	&	0.28	&	-	&	-	&	1.5	&	-	&	5.3	\\
0329-071	&	0.29	&	-	&	-	&	2.3	&	-	&	8.1	\\
0331-013	&	0.3	&	0.78	&	0.83	&	13.5	&	38	&	45.6	\\
0333+130	&	0.29	&	-	&	-	&	2.9	&	-	&	9.9	\\
0333+095	&	0.11	&	-	&	-	&	1.2	&	-	&	10.8	\\
0334+340	&	0.3	&	0.89	&	1	&	1.6	&	5.4	&	5.4	\\
0335+125	&	0.91	&	0.24	&	1	&	4.9	&	5.4	&	5.4	\\
0335+100	&	0.12	&	-	&	-	&	0.9	&	-	&	7.6	\\
0335-006	&	0.36	&	-	&	-	&	1.9	&	-	&	5.3	\\
0336+096	&	0.11	&	-	&	-	&	0.7	&	-	&	5.9	\\
0341+092	&	0.28	&	-	&	-	&	4.2	&	-	&	15.1	\\
0341+064	&	0.16	&	0.67	&	0.93	&	0.9	&	5.2	&	5.6	\\
0343+277	&	0.14	&	-	&	-	&	1.2	&	-	&	8.5	\\
0343+337	&	0.27	&	0.67	&	0.96	&	2.6	&	9	&	9.4	\\
0345+340	&	0.24	&	0.58	&	0.68	&	3.5	&	9.8	&	14.4	\\
0346+301	&	0.24	&	-	&	-	&	1.4	&	-	&	5.6	\\
0346+335	&	0.3	&	-	&	-	&	3.9	&	-	&	13.2	\\
0347+055	&	0.21	&	0.83	&	0.76	&	6.9	&	25.3	&	33.4	\\
0348+263	&	0.06	&	-	&	-	&	0.8	&	-	&	12.8	\\
0349+126	&	0.59	&	-	&	-	&	3.1	&	-	&	5.3	\\
0349+383	&	0.16	&	-	&	-	&	1.5	&	-	&	9.7	\\
0350-072	&	0.22	&	-	&	-	&	6.2	&	-	&	28.1	\\
0352-081	&	0.16	&	-	&	-	&	1.4	&	-	&	8.7	\\
0353+129	&	0.27	&	0.74	&	0.9	&	4.2	&	13.8	&	15.3	\\
0355+217	&	0.14	&	-	&	-	&	0.7	&	-	&	5.2	\\
0355+145	&	0.21	&	-	&	-	&	1.9	&	-	&	9.1	\\
0356+237	&	0.17	&	-	&	-	&	0.9	&	-	&	5.3	\\
0401+159	&	0.42	&	0.38	&	0.75	&	2.2	&	3.9	&	5.2	\\
0403+303	&	0.09	&	-	&	-	&	2.2	&	-	&	24.1	\\
0405+259	&	0.42	&	-	&	-	&	2.2	&	-	&	5.2	\\
0406+387	&	0.26	&	-	&	-	&	2.7	&	-	&	10.3	\\
0406+294	&	0.24	&	-	&	-	&	1.5	&	-	&	6.4	\\
0408+070	&	0.25	&	-	&	-	&	1.3	&	-	&	5.1	\\
0408+170	&	0.15	&	-	&	-	&	1.4	&	-	&	9.3	\\
0409+229	&	0.31	&	-	&	-	&	2.6	&	-	&	8.5	\\
0410+267	&	0.59	&	-	&	-	&	5.4	&	-	&	9.1	\\
0410+111	&	0.07	&	-	&	-	&	2.4	&	-	&	33.9	\\
0417+253	&	0.19	&	-	&	-	&	2.5	&	-	&	13.2	\\
0418+106	&	0.17	&	-	&	-	&	1	&	-	&	5.9	\\
0418+151	&	0.11	&	-	&	-	&	1.5	&	-	&	13.4	\\
0418+146	&	0.12	&	-	&	-	&	0.9	&	-	&	7.2	\\
0420+349	&	0.19	&	-	&	-	&	1.7	&	-	&	8.7	\\
0423+307	&	0.19	&	-	&	-	&	1.5	&	-	&	7.8	\\
0425+233	&	0.44	&	-	&	-	&	4.3	&	-	&	9.9	\\
0426+411	&	0.28	&	0.37	&	1	&	1.7	&	6	&	6	\\
0426+157	&	0.12	&	-	&	-	&	0.6	&	-	&	5.3	\\
0430+050	&	0.31	&	0.41	&	0.58	&	2.8	&	5.2	&	8.9	\\
0433+032	&	0.13	&	-	&	-	&	1	&	-	&	7.9	\\
0433+213	&	0.13	&	-	&	-	&	2.5	&	-	&	19.1	\\
0434+102	&	0.37	&	-	&	-	&	1.8	&	-	&	4.9	\\
0435+230	&	0.12	&	-	&	-	&	1.8	&	-	&	15.4	\\
0439+013	&	0.19	&	0.42	&	0.29	&	3.7	&	5.5	&	19.2	\\
0442+151	&	0.07	&	-	&	-	&	0.5	&	-	&	7.6	\\
0442+396	&	0.16	&	-	&	-	&	4.5	&	-	&	28	\\
0444+334	&	0.08	&	-	&	-	&	1.3	&	-	&	15.8	\\
0446+207	&	0.59	&	-	&	-	&	2.8	&	-	&	4.7	\\
0447+014	&	0.21	&	-	&	-	&	1.6	&	-	&	7.6	\\
0449+319	&	0.12	&	-	&	-	&	0.9	&	-	&	7.6	\\
0450+314	&	0.09	&	-	&	-	&	2.3	&	-	&	25.8	\\
0452+003	&	0.28	&	0.66	&	0.72	&	2.2	&	5.9	&	8.2	\\
0453+141	&	0.43	&	0.57	&	1	&	3.7	&	8.6	&	8.6	\\
0453+210	&	0.15	&	-	&	-	&	0.8	&	-	&	5.4	\\
0456+324	&	0.15	&	-	&	-	&	1	&	-	&	6.5	\\
0456+397	&	0.11	&	-	&	-	&	0.7	&	-	&	6.2	\\
0459+247	&	0.2	&	-	&	-	&	1.1	&	-	&	5.4	\\
0459+252	&	0.13	&	-	&	-	&	4.8	&	-	&	36.9	\\
0502+282	&	0.11	&	-	&	-	&	1.8	&	-	&	16	\\
0505+034	&	0.17	&	-	&	-	&	0.8	&	-	&	4.8	\\
0508+028	&	0.14	&	-	&	-	&	2.8	&	-	&	20.2	\\
0510+387	&	0.37	&	-	&	-	&	5.8	&	-	&	15.7	\\
0511+017	&	0.18	&	0.74	&	0.44	&	3.4	&	8.4	&	18.9	\\
0511+337	&	0.07	&	-	&	-	&	1	&	-	&	14.9	\\
0511+347	&	0.09	&	-	&	-	&	0.6	&	-	&	6.5	\\
0512-014	&	0.37	&	0.74	&	1	&	2.3	&	6.2	&	6.2	\\
0513+249	&	0.09	&	-	&	-	&	2.4	&	-	&	26.8	\\
0514+266	&	0.22	&	-	&	-	&	1.2	&	-	&	5.3	\\
0514+238	&	0.29	&	-	&	-	&	2.1	&	-	&	7.2	\\
0514+244	&	0.14	&	-	&	-	&	0.8	&	-	&	5.4	\\
0514+107	&	0.13	&	-	&	-	&	1.2	&	-	&	9.4	\\
0518+165	&	0.56	&	0.38	&	0.98	&	16.6	&	28.6	&	29.3	\\
0519+146	&	0.1	&	-	&	-	&	1	&	-	&	9.9	\\
0519-060	&	0.26	&	-	&	-	&	1.2	&	-	&	4.7	\\
0523+328	&	0.09	&	-	&	-	&	2.7	&	-	&	30.4	\\
0537+174	&	0.22	&	-	&	-	&	1.1	&	-	&	4.8	\\
0542-010	&	0.3	&	-	&	-	&	1.7	&	-	&	5.8	\\
0544+011	&	0.17	&	0.62	&	0.51	&	1.6	&	4.8	&	9.4	\\
0549+085	&	0.3	&	-	&	-	&	1.6	&	-	&	5.3	\\
0552+066	&	0.16	&	-	&	-	&	0.9	&	-	&	5.3	\\
0556+283	&	0.61	&	-	&	-	&	3.4	&	-	&	5.5	\\
0558+387	&	0.26	&	-	&	-	&	2.5	&	-	&	9.7	\\
0601+299	&	0.1	&	-	&	-	&	1.1	&	-	&	11.3	\\
0601+012	&	0.24	&	-	&	-	&	1.1	&	-	&	4.8	\\
0602+018	&	0.18	&	-	&	-	&	1	&	-	&	5.4	\\
0606+124	&	0.13	&	-	&	-	&	0.8	&	-	&	6	\\
0607+366	&	0.43	&	-	&	-	&	2.3	&	-	&	5.4	\\
0607+094	&	0.33	&	-	&	-	&	1.7	&	-	&	5.3	\\
0610+261	&	0.1	&	-	&	-	&	2.9	&	-	&	28.7	\\
0612+415	&	0.36	&	-	&	-	&	3.1	&	-	&	8.6	\\
0614+196	&	0.57	&	-	&	-	&	11.9	&	-	&	20.9	\\
0614+139	&	0.24	&	-	&	-	&	5.5	&	-	&	22.8	\\
0614+376	&	0.07	&	-	&	-	&	0.9	&	-	&	13.2	\\
0616+341	&	0.14	&	-	&	-	&	1.1	&	-	&	7.9	\\
0618+268	&	0.14	&	-	&	-	&	0.8	&	-	&	5.5	\\
0619+145	&	0.24	&	-	&	-	&	6.1	&	-	&	25.4	\\
0619+230	&	0.11	&	-	&	-	&	1.1	&	-	&	10	\\
0619+384	&	0.13	&	-	&	-	&	2.4	&	-	&	18.3	\\
0620-026	&	0.34	&	-	&	-	&	3.6	&	-	&	10.6	\\
0620+318	&	0.39	&	-	&	-	&	5.3	&	-	&	13.5	\\
0621+402	&	0.12	&	-	&	-	&	4.5	&	-	&	37.9	\\
0625+337	&	0.17	&	-	&	-	&	1.6	&	-	&	9.4	\\
0627-014	&	0.25	&	-	&	-	&	1.6	&	-	&	6.6	\\
0627+003	&	0.23	&	0.55	&	0.52	&	2	&	4.7	&	9	\\
0628+251	&	0.06	&	-	&	-	&	1	&	-	&	15.9	\\
0630-030	&	0.29	&	-	&	-	&	2	&	-	&	6.8	\\
0631+117	&	0.59	&	-	&	-	&	4.9	&	-	&	8.3	\\
0632+023	&	0.12	&	-	&	-	&	4.4	&	-	&	36.8	\\
0633+111	&	0.37	&	-	&	-	&	2.5	&	-	&	6.8	\\
0633-027	&	0.31	&	-	&	-	&	2.5	&	-	&	8.2	\\
0635-002	&	0.34	&	-	&	-	&	1.8	&	-	&	5.3	\\
0641+074	&	0.16	&	-	&	-	&	0.8	&	-	&	4.7	\\
0644+419	&	0.15	&	-	&	-	&	1	&	-	&	6.9	\\
0644+191	&	0.08	&	-	&	-	&	0.5	&	-	&	6.5	\\
0645-021	&	0.34	&	-	&	-	&	7	&	-	&	20.5	\\
0646-083	&	0.28	&	-	&	-	&	1.5	&	-	&	5.3	\\
0648+376	&	0.24	&	-	&	-	&	1.8	&	-	&	7.3	\\
0649+227	&	0.12	&	-	&	-	&	3.1	&	-	&	26.2	\\
0651+240	&	0.21	&	-	&	-	&	1.2	&	-	&	5.7	\\
0653+106	&	0.21	&	-	&	-	&	2.1	&	-	&	9.9	\\
0655+360	&	0.23	&	-	&	-	&	2.7	&	-	&	11.8	\\
0655+169	&	0.22	&	-	&	-	&	1.3	&	-	&	5.8	\\
0656+214	&	0.86	&	-	&	-	&	4.9	&	-	&	5.7	\\
0657+325	&	0.3	&	-	&	-	&	1.5	&	-	&	5	\\
0657+032	&	0.27	&	-	&	-	&	1.3	&	-	&	4.8	\\
0658+354	&	0.31	&	0.69	&	0.86	&	5.1	&	14.3	&	16.5	\\
0658+380	&	0.33	&	0.53	&	0.8	&	7	&	16.8	&	21.2	\\
0659-024	&	0.29	&	-	&	-	&	7.8	&	-	&	27	\\
0700+375	&	0.35	&	0.35	&	0.57	&	2.6	&	4.2	&	7.3	\\
0704+351	&	0.13	&	-	&	-	&	0.6	&	-	&	4.7	\\
0709-088	&	0.29	&	-	&	-	&	2.8	&	-	&	9.6	\\
0710+118	&	0.08	&	-	&	-	&	2	&	-	&	25.4	\\
0711+147	&	0.25	&	-	&	-	&	3.3	&	-	&	13.3	\\
0712-089	&	0.34	&	0.82	&	1	&	1.8	&	5.3	&	5.3	\\
0713+367	&	0.09	&	-	&	-	&	0.4	&	-	&	4.9	\\
0715+378	&	0.07	&	-	&	-	&	0.9	&	-	&	12.9	\\
0720+192	&	0.15	&	-	&	-	&	1.5	&	-	&	10.2	\\
0721+129	&	0.28	&	-	&	-	&	2.8	&	-	&	10.1	\\
0721+161	&	0.13	&	-	&	-	&	1.6	&	-	&	12.4	\\
0725+146	&	0.35	&	0.62	&	0.96	&	5.7	&	15.5	&	16.2	\\
0727+214	&	0.37	&	-	&	-	&	2.5	&	-	&	6.9	\\
0730+259	&	0.18	&	-	&	-	&	1	&	-	&	5.5	\\
0731+270	&	0.49	&	-	&	-	&	2.6	&	-	&	5.3	\\
0731+317	&	0.24	&	-	&	-	&	2.1	&	-	&	8.7	\\
0732+332	&	0.59	&	-	&	-	&	8.1	&	-	&	13.8	\\
0732+291	&	0.13	&	-	&	-	&	1.6	&	-	&	11.9	\\
0733+244	&	0.13	&	0.69	&	0.41	&	0.6	&	2	&	4.8	\\
0733+361	&	0.39	&	-	&	-	&	3.7	&	-	&	9.5	\\
0740+380	&	0.3	&	-	&	-	&	9.6	&	-	&	32	\\
0741+394	&	0.21	&	-	&	-	&	2	&	-	&	9.4	\\
0742+376	&	0.22	&	-	&	-	&	1.9	&	-	&	8.7	\\
0745+343	&	0.21	&	-	&	-	&	1	&	-	&	5	\\
0745+119	&	0.48	&	0.32	&	0.72	&	2.7	&	4.1	&	5.6	\\
0747+316	&	0.38	&	-	&	-	&	3.5	&	-	&	9.2	\\
0748+343	&	0.19	&	0.6	&	0.62	&	1.8	&	6	&	9.7	\\
0750+299	&	0.18	&	-	&	-	&	1.3	&	-	&	7.1	\\
0755+379	&	0.08	&	-	&	-	&	1.5	&	-	&	18.6	\\
0800-040	&	0.44	&	-	&	-	&	2.8	&	-	&	6.4	\\
0801+304	&	0.3	&	-	&	-	&	1.9	&	-	&	6.2	\\
0807+285	&	0.19	&	-	&	-	&	0.9	&	-	&	4.8	\\
0810+371	&	0.28	&	-	&	-	&	2.1	&	-	&	7.4	\\
0811+389	&	0.16	&	-	&	-	&	1.7	&	-	&	10.4	\\
0812+381	&	0.12	&	-	&	-	&	0.9	&	-	&	7.4	\\
0820+225	&	0.2	&	-	&	-	&	1.5	&	-	&	7.3	\\
0822+345	&	0.09	&	-	&	-	&	1.1	&	-	&	12.7	\\
0823+379	&	0.46	&	0.24	&	0.8	&	3.4	&	6	&	7.4	\\
0827+379	&	0.6	&	0.28	&	0.85	&	8	&	11.4	&	13.3	\\
0831-052	&	0.27	&	-	&	-	&	3.4	&	-	&	12.7	\\
0831+099	&	0.68	&	-	&	-	&	3.6	&	-	&	5.3	\\
0834+371	&	0.11	&	-	&	-	&	0.5	&	-	&	4.8	\\
0836+255	&	0.1	&	-	&	-	&	2	&	-	&	19.6	\\
0837+300	&	0.09	&	-	&	-	&	0.4	&	-	&	4.8	\\
0846+380	&	0.06	&	-	&	-	&	0.4	&	-	&	6.9	\\
0850+344	&	0.23	&	-	&	-	&	2.3	&	-	&	10.1	\\
0854+343	&	0.29	&	-	&	-	&	4.9	&	-	&	16.9	\\
0855+281	&	0.2	&	-	&	-	&	3.5	&	-	&	17.7	\\
0855-039	&	0.31	&	-	&	-	&	1.5	&	-	&	5	\\
0857+188	&	0.08	&	-	&	-	&	0.8	&	-	&	9.9	\\
0858+388	&	0.12	&	-	&	-	&	2.2	&	-	&	18.6	\\
0902+225	&	0.06	&	-	&	-	&	0.6	&	-	&	10.1	\\
0903+258	&	0.11	&	-	&	-	&	0.9	&	-	&	8.1	\\
0906+381	&	0.09	&	-	&	-	&	3.3	&	-	&	36.5	\\
0915+053	&	0.22	&	-	&	-	&	1.1	&	-	&	5	\\
0916+019	&	0.36	&	0.62	&	0.65	&	1.7	&	3	&	4.7	\\
0918+219	&	0.39	&	-	&	-	&	2.2	&	-	&	5.5	\\
0919+314	&	0.41	&	0.43	&	0.79	&	10	&	19.4	&	24.6	\\
0923+392	&	0.19	&	-	&	-	&	1.3	&	-	&	6.9	\\
0926+280	&	0.28	&	-	&	-	&	1.4	&	-	&	5.1	\\
0927+315	&	0.11	&	-	&	-	&	0.7	&	-	&	6.4	\\
0927+362	&	0.23	&	-	&	-	&	3.3	&	-	&	14.3	\\
0932+399	&	0.39	&	-	&	-	&	5.8	&	-	&	14.9	\\
0939+267	&	0.54	&	0.69	&	1	&	3.6	&	6.7	&	6.7	\\
0947+145	&	0.11	&	-	&	-	&	2.3	&	-	&	21.3	\\
0949+003	&	0.32	&	0.53	&	0.86	&	7.2	&	19.2	&	22.2	\\
0951+216	&	0.08	&	-	&	-	&	0.7	&	-	&	8.8	\\
0951+377	&	0.19	&	-	&	-	&	1.1	&	-	&	5.6	\\
0952+358	&	0.16	&	-	&	-	&	1.9	&	-	&	11.8	\\
0954+278	&	0.1	&	-	&	-	&	0.7	&	-	&	6.9	\\
0956+389	&	0.12	&	-	&	-	&	1	&	-	&	8.3	\\
0959+290	&	0.07	&	-	&	-	&	3.9	&	-	&	56.1	\\
1001+321	&	0.2	&	-	&	-	&	2.8	&	-	&	14	\\
1001+106	&	0.32	&	0.56	&	0.82	&	2.8	&	7.3	&	8.8	\\
1002+353	&	0.22	&	0.5	&	0.71	&	4.6	&	14.6	&	20.6	\\
1003+147	&	0.13	&	-	&	-	&	0.6	&	-	&	5	\\
1005+121	&	0.13	&	-	&	-	&	0.6	&	-	&	4.7	\\
1005+078	&	0.28	&	0.53	&	0.73	&	12.7	&	32.6	&	44.8	\\
1008+067	&	0.21	&	0.58	&	0.52	&	8.4	&	21	&	40.3	\\
1008+215	&	0.25	&	-	&	-	&	2	&	-	&	7.8	\\
1008+322	&	0.13	&	-	&	-	&	0.9	&	-	&	7.2	\\
1010+234	&	0.18	&	-	&	-	&	1	&	-	&	5.3	\\
1010+408	&	0.31	&	-	&	-	&	2.8	&	-	&	8.9	\\
1011+292	&	0.17	&	-	&	-	&	1.9	&	-	&	11.5	\\
1013+129	&	0.26	&	0.78	&	1	&	1.4	&	5.4	&	5.4	\\
1014+393	&	0.27	&	-	&	-	&	1.5	&	-	&	5.6	\\
1014+277	&	0.28	&	0.54	&	0.58	&	3.3	&	7	&	12.1	\\
1019+223	&	0.25	&	0.56	&	0.57	&	5.5	&	12.6	&	22.1	\\
1023+068	&	0.13	&	-	&	-	&	4	&	-	&	30.6	\\
1025+154	&	0.65	&	0.24	&	0.88	&	4.1	&	5.6	&	6.4	\\
1026+391	&	0.18	&	-	&	-	&	1.1	&	-	&	6.2	\\
1029+253	&	0.1	&	-	&	-	&	0.7	&	-	&	6.6	\\
1033+264	&	0.11	&	-	&	-	&	0.6	&	-	&	5.1	\\
1035+363	&	0.15	&	-	&	-	&	1.2	&	-	&	8.2	\\
1036+323	&	0.11	&	-	&	-	&	0.6	&	-	&	5.4	\\
1036-043	&	0.42	&	0.65	&	1	&	2.7	&	6.3	&	6.3	\\
1039+030	&	0.17	&	0.86	&	0.41	&	3.2	&	8	&	19.4	\\
1040+062	&	0.29	&	0.48	&	0.66	&	4.2	&	9.7	&	14.6	\\
1042+112	&	0.31	&	0.62	&	0.82	&	1.6	&	4.1	&	5.1	\\
1042+392	&	0.1	&	-	&	-	&	1.1	&	-	&	11.4	\\
1044+227	&	0.08	&	-	&	-	&	0.4	&	-	&	5.5	\\
1044+298	&	0.28	&	-	&	-	&	1.6	&	-	&	5.8	\\
1046+358	&	0.12	&	-	&	-	&	1.1	&	-	&	8.9	\\
1047+288	&	0.25	&	-	&	-	&	2	&	-	&	8.1	\\
1048+098	&	0.14	&	-	&	-	&	0.7	&	-	&	5.3	\\
1049+044	&	0.22	&	1	&	0.94	&	1	&	4.5	&	4.8	\\
1055+316	&	0.23	&	-	&	-	&	1.2	&	-	&	5.3	\\
1057+307	&	0.1	&	-	&	-	&	0.9	&	-	&	8.6	\\
1059+101	&	0.41	&	0.55	&	1	&	2	&	4.8	&	4.8	\\
1059-009	&	0.35	&	-	&	-	&	5.2	&	-	&	14.9	\\
1059+031	&	0.2	&	0.6	&	0.72	&	1.1	&	4	&	5.5	\\
1059+108	&	0.2	&	0.63	&	0.54	&	1.9	&	5.3	&	9.9	\\
1104+160	&	0.13	&	-	&	-	&	0.6	&	-	&	4.7	\\
1104+129	&	0.16	&	-	&	-	&	1	&	-	&	6.2	\\
1104+314	&	0.15	&	-	&	-	&	1	&	-	&	6.5	\\
1105+392	&	0.11	&	-	&	-	&	1.1	&	-	&	9.7	\\
1106+253	&	0.08	&	0.7	&	0	&	2.4	&	0.1	&	31.3	\\
1107+379	&	0.09	&	-	&	-	&	1.6	&	-	&	18.2	\\
1107+043	&	0.35	&	0.48	&	0.81	&	2.4	&	5.6	&	6.9	\\
1108+033	&	0.21	&	0.78	&	0.64	&	2.6	&	8.2	&	12.8	\\
1108+360	&	0.1	&	-	&	-	&	2.8	&	-	&	27.6	\\
1109+411	&	0.14	&	-	&	-	&	1.8	&	-	&	12.6	\\
1111+409	&	0.11	&	-	&	-	&	6	&	-	&	54.6	\\
1113+218	&	0.15	&	-	&	-	&	1.9	&	-	&	12.9	\\
1113+273	&	0.11	&	-	&	-	&	1.1	&	-	&	9.8	\\
1118+238	&	0.46	&	-	&	-	&	5.4	&	-	&	11.6	\\
1119+127	&	0.2	&	-	&	-	&	7	&	-	&	35	\\
1119+197	&	0.71	&	0.77	&	1	&	4.9	&	7	&	7	\\
1120+217	&	0.12	&	-	&	-	&	1.2	&	-	&	10.1	\\
1123+303	&	0.13	&	-	&	-	&	3.4	&	-	&	26	\\
1124+260	&	0.13	&	-	&	-	&	0.9	&	-	&	7.2	\\
1126+018	&	0.19	&	-	&	-	&	3	&	-	&	15.8	\\
1130+106	&	0.47	&	-	&	-	&	2.3	&	-	&	4.8	\\
1132+304	&	0.1	&	-	&	-	&	0.9	&	-	&	8.6	\\
1133+262	&	0.11	&	-	&	-	&	0.6	&	-	&	5.8	\\
1135+314	&	0.08	&	-	&	-	&	0.6	&	-	&	7.5	\\
1138+123	&	0.08	&	-	&	-	&	0.7	&	-	&	9.1	\\
1138+058	&	0.12	&	-	&	-	&	0.7	&	-	&	5.7	\\
1139+063	&	0.34	&	0.79	&	1	&	1.6	&	4.7	&	4.7	\\
1139+234	&	0.12	&	-	&	-	&	1.5	&	-	&	12.5	\\
1140+218	&	0.11	&	-	&	-	&	1.9	&	-	&	17.1	\\
1140+224	&	0.25	&	-	&	-	&	6.4	&	-	&	25.8	\\
1141+303	&	0.11	&	-	&	-	&	1.5	&	-	&	13.8	\\
1141+374	&	0.34	&	0.41	&	0.63	&	4	&	7.3	&	11.6	\\
1142+319	&	0.09	&	-	&	-	&	3.8	&	-	&	41.9	\\
1142+354	&	0.12	&	-	&	-	&	1.1	&	-	&	8.9	\\
1143+293	&	0.1	&	-	&	-	&	1.2	&	-	&	12	\\
1145+256	&	0.15	&	-	&	-	&	0.9	&	-	&	6.2	\\
1147+131	&	0.25	&	0.63	&	0.63	&	4.4	&	11.2	&	17.9	\\
1148+287	&	0.08	&	-	&	-	&	0.4	&	-	&	5.5	\\
1148+387	&	0.14	&	-	&	-	&	1	&	-	&	7.4	\\
1149+174	&	0.16	&	-	&	-	&	0.9	&	-	&	5.3	\\
1150+334	&	0.11	&	-	&	-	&	0.6	&	-	&	5.1	\\
1150+115	&	0.21	&	-	&	-	&	1.6	&	-	&	7.7	\\
1151+026	&	0.2	&	0.96	&	0.69	&	1.1	&	3.7	&	5.4	\\
1152+045	&	0.18	&	0.86	&	0.59	&	3.3	&	11	&	18.8	\\
1153+305	&	0.23	&	0.92	&	0.91	&	1.2	&	4.6	&	5.1	\\
1154+193	&	0.19	&	-	&	-	&	0.9	&	-	&	5	\\
1157+255	&	0.11	&	-	&	-	&	1	&	-	&	9.2	\\
1158+123	&	0.19	&	0.31	&	1	&	1	&	5.4	&	5.4	\\
1159-060	&	0.3	&	-	&	-	&	1.6	&	-	&	5.4	\\
1200+353	&	0.12	&	0.65	&	0.86	&	0.9	&	6.2	&	7.2	\\
1201+240	&	0.08	&	-	&	-	&	0.4	&	-	&	4.7	\\
1201+310	&	0.13	&	-	&	-	&	0.8	&	-	&	6	\\
1204+372	&	0.16	&	0.94	&	0.77	&	0.9	&	4.2	&	5.5	\\
1204+353	&	0.22	&	0.92	&	0.87	&	2.1	&	8.2	&	9.5	\\
1209+120	&	0.24	&	-	&	-	&	1.4	&	-	&	5.9	\\
1211+285	&	0.13	&	-	&	-	&	1	&	-	&	7.5	\\
1211-007	&	0.46	&	0.48	&	0.91	&	3	&	6.1	&	6.7	\\
1212+261	&	0.14	&	-	&	-	&	0.8	&	-	&	5.4	\\
1214+351	&	0.15	&	-	&	-	&	1.5	&	-	&	10.2	\\
1216-068	&	0.48	&	0.96	&	1	&	2.4	&	5	&	5	\\
1216-021	&	0.34	&	0.68	&	0.93	&	7.2	&	19.8	&	21.3	\\
1216+235	&	0.13	&	-	&	-	&	0.7	&	-	&	5.3	\\
1216+194	&	0.21	&	-	&	-	&	1	&	-	&	4.7	\\
1217+001	&	0.34	&	0.46	&	0.69	&	2.3	&	4.7	&	6.8	\\
1217-038	&	0.36	&	0.67	&	1	&	2.1	&	5.8	&	5.8	\\
1218+340	&	0.16	&	0.95	&	0.79	&	3.8	&	18.5	&	23.3	\\
1218-026	&	0.35	&	0.89	&	1	&	1.8	&	5.2	&	5.2	\\
1218+228	&	0.21	&	-	&	-	&	2.1	&	-	&	9.9	\\
1218+318	&	0.09	&	-	&	-	&	0.9	&	-	&	10.5	\\
1222+217	&	0.14	&	-	&	-	&	1.3	&	-	&	9.1	\\
1225+207	&	0.22	&	-	&	-	&	1.6	&	-	&	7.1	\\
1228+263	&	0.12	&	-	&	-	&	0.7	&	-	&	5.7	\\
1228+298	&	0.29	&	0.57	&	0.76	&	1.5	&	4	&	5.3	\\
1228+419	&	0.37	&	0.72	&	0.82	&	2.7	&	6	&	7.3	\\
1229+342	&	0.18	&	-	&	-	&	0.9	&	-	&	5.1	\\
1232+397	&	0.16	&	0.87	&	0.58	&	1.9	&	6.9	&	12	\\
1233+268	&	0.07	&	-	&	-	&	0.4	&	-	&	6.2	\\
1233+168	&	0.1	&	0.48	&	0	&	1.4	&	0	&	13.3	\\
1234+372	&	0.15	&	-	&	-	&	1.2	&	-	&	8.2	\\
1236+057	&	0.12	&	-	&	-	&	2.7	&	-	&	22.6	\\
1239+329	&	0.13	&	-	&	-	&	2.1	&	-	&	16.5	\\
1242+364	&	0.55	&	0.3	&	0.85	&	4.2	&	6.4	&	7.5	\\
1244+390	&	0.13	&	-	&	-	&	1.6	&	-	&	12.5	\\
1246+095	&	0.26	&	-	&	-	&	3.4	&	-	&	13.3	\\
1246+220	&	0.2	&	-	&	-	&	1	&	-	&	5.1	\\
1246+113	&	0.23	&	-	&	-	&	1.1	&	-	&	4.8	\\
1248+240	&	0.11	&	-	&	-	&	0.5	&	-	&	4.9	\\
1251+160	&	0.12	&	-	&	-	&	1.8	&	-	&	15	\\
1251+349	&	0.16	&	-	&	-	&	0.8	&	-	&	4.8	\\
1252+278	&	0.06	&	-	&	-	&	1.4	&	-	&	23.3	\\
1253+375	&	0.09	&	-	&	-	&	0.9	&	-	&	10	\\
1255+370	&	0.14	&	-	&	-	&	1.8	&	-	&	13.1	\\
1255-078	&	0.4	&	-	&	-	&	2.3	&	-	&	5.7	\\
1257-003	&	0.5	&	0.55	&	0.99	&	3	&	5.9	&	5.9	\\
1258+384	&	0.11	&	-	&	-	&	0.7	&	-	&	6.5	\\
1258+404	&	0.14	&	-	&	-	&	3.5	&	-	&	25.1	\\
1258+300	&	0.14	&	-	&	-	&	1	&	-	&	7.2	\\
1302+388	&	0.21	&	-	&	-	&	1.7	&	-	&	8.3	\\
1303+091	&	0.19	&	-	&	-	&	4.1	&	-	&	21.5	\\
1304+067	&	0.08	&	-	&	-	&	0.4	&	-	&	4.8	\\
1307+122	&	0.36	&	0.9	&	1	&	4.4	&	12.3	&	12.3	\\
1309+212	&	0.65	&	0.78	&	1	&	3.5	&	5.4	&	5.4	\\
1313+022	&	0.36	&	0.49	&	0.62	&	10.1	&	17.3	&	27.9	\\
1315+233	&	0.23	&	-	&	-	&	1.6	&	-	&	7.1	\\
1315+349	&	0.18	&	-	&	-	&	0.9	&	-	&	5.1	\\
1316+299	&	0.21	&	0.71	&	1	&	2.2	&	10.5	&	10.5	\\
1318+258	&	0.24	&	-	&	-	&	1.1	&	-	&	4.7	\\
1318+380	&	0.18	&	-	&	-	&	1.8	&	-	&	10	\\
1318+114	&	0.3	&	-	&	-	&	4.3	&	-	&	14.2	\\
1319+271	&	0.47	&	-	&	-	&	3.6	&	-	&	7.7	\\
1320+299	&	0.21	&	-	&	-	&	2	&	-	&	9.4	\\
1321+318	&	0.09	&	-	&	-	&	0.7	&	-	&	7.7	\\
1322+259	&	0.22	&	-	&	-	&	1.1	&	-	&	4.8	\\
1324+230	&	0.42	&	-	&	-	&	2.8	&	-	&	6.6	\\
1325+371	&	0.08	&	-	&	-	&	0.5	&	-	&	6.1	\\
1325+321	&	0.72	&	-	&	-	&	4.7	&	-	&	6.5	\\
1326+150	&	0.07	&	-	&	-	&	0.8	&	-	&	12	\\
1326+070	&	0.32	&	-	&	-	&	2	&	-	&	6.2	\\
1326+312	&	0.32	&	0.39	&	0.64	&	2.9	&	5.8	&	9.1	\\
1328+308	&	0.3	&	0.52	&	0.64	&	7.8	&	16.7	&	26.3	\\
1328+253	&	0.54	&	0.36	&	0.93	&	16.2	&	28	&	30.1	\\
1329+140	&	0.14	&	-	&	-	&	1.4	&	-	&	9.8	\\
1336+359	&	0.15	&	-	&	-	&	1.2	&	-	&	7.9	\\
1340+354	&	0.1	&	-	&	-	&	0.7	&	-	&	6.9	\\
1340+088	&	0.45	&	-	&	-	&	6.6	&	-	&	14.7	\\
1340+320	&	0.08	&	0.77	&	0	&	0.5	&	0	&	6.4	\\
1341+144	&	0.46	&	0.38	&	0.85	&	4.1	&	7.6	&	8.9	\\
1343+215	&	0.13	&	-	&	-	&	0.9	&	-	&	7.2	\\
1345+288	&	0.11	&	-	&	-	&	1.1	&	-	&	9.8	\\
1345+245	&	0.32	&	-	&	-	&	2.7	&	-	&	8.4	\\
1346+269	&	0.09	&	-	&	-	&	0.6	&	-	&	6.8	\\
1347+215	&	0.19	&	-	&	-	&	3	&	-	&	15.6	\\
1348+161	&	0.18	&	-	&	-	&	0.9	&	-	&	4.8	\\
1349+353	&	0.24	&	-	&	-	&	1.5	&	-	&	6.4	\\
1350+317	&	0.11	&	-	&	-	&	2.5	&	-	&	22.5	\\
1352+164	&	0.06	&	-	&	-	&	0.9	&	-	&	15.5	\\
1353+259	&	0.11	&	-	&	-	&	0.6	&	-	&	5.7	\\
1354+398	&	0.18	&	-	&	-	&	1.7	&	-	&	9.4	\\
1357+267	&	0.12	&	-	&	-	&	0.7	&	-	&	5.4	\\
1358+116	&	0.25	&	-	&	-	&	1.3	&	-	&	5.3	\\
1401+387	&	0.09	&	-	&	-	&	0.8	&	-	&	8.6	\\
1401+353	&	0.24	&	-	&	-	&	2.6	&	-	&	11	\\
1403-026	&	0.63	&	0.47	&	1	&	4.6	&	7.2	&	7.2	\\
1404+344	&	0.12	&	-	&	-	&	3.2	&	-	&	26.4	\\
1405+257	&	0.28	&	0.5	&	0.68	&	1.4	&	3.3	&	4.9	\\
1405+241	&	0.3	&	-	&	-	&	2.9	&	-	&	9.6	\\
1407+317	&	0.08	&	-	&	-	&	1.4	&	-	&	17	\\
1408+370	&	0.37	&	0.47	&	0.79	&	3.1	&	6.6	&	8.4	\\
1414+111	&	0.07	&	-	&	-	&	1.2	&	-	&	16.8	\\
1414+362	&	0.1	&	-	&	-	&	0.5	&	-	&	5.3	\\
1414-042	&	0.25	&	0.65	&	0.75	&	2.6	&	7.9	&	10.5	\\
1415+072	&	0.26	&	0.87	&	0.88	&	6.2	&	21.4	&	24.2	\\
1416+045	&	0.15	&	0.91	&	0.55	&	3.2	&	11.6	&	21	\\
1416+059	&	0.36	&	0.55	&	0.91	&	7.7	&	19.3	&	21.3	\\
1416+142	&	0.43	&	0.34	&	0.7	&	2.1	&	3.5	&	5	\\
1416+067	&	0.34	&	0.52	&	0.81	&	21.3	&	50.1	&	62	\\
1417+272	&	0.12	&	-	&	-	&	0.8	&	-	&	6.9	\\
1423+243	&	0.16	&	-	&	-	&	1.7	&	-	&	10.6	\\
1423+275	&	0.33	&	0.76	&	0.71	&	3.7	&	7.8	&	11	\\
1424+008	&	0.44	&	0.34	&	0.49	&	2.2	&	2.5	&	5.1	\\
1425+287	&	0.46	&	-	&	-	&	3.5	&	-	&	7.7	\\
1425+044	&	0.3	&	0.94	&	1	&	2.1	&	7.1	&	7.1	\\
1426+296	&	0.23	&	0.99	&	0.78	&	1.3	&	4.5	&	5.8	\\
1427+075	&	0.13	&	-	&	-	&	1.6	&	-	&	12.6	\\
1435+038	&	0.11	&	-	&	-	&	4.9	&	-	&	44.3	\\
1436+286	&	0.34	&	0.5	&	0.62	&	2.3	&	4.2	&	6.8	\\
1437-076	&	0.22	&	0.89	&	0.93	&	3.3	&	13.9	&	14.9	\\
1438+358	&	0.19	&	-	&	-	&	1.7	&	-	&	8.9	\\
1443+240	&	0.2	&	-	&	-	&	1.2	&	-	&	5.8	\\
1445+376	&	0.32	&	0.6	&	0.97	&	1.9	&	5.7	&	5.9	\\
1454+244	&	0.18	&	0.88	&	0.68	&	1.1	&	4.1	&	6.1	\\
1454+271	&	0.22	&	-	&	-	&	1.4	&	-	&	6.3	\\
1455+253	&	0.07	&	-	&	-	&	0.6	&	-	&	8.6	\\
1457+146	&	0.12	&	-	&	-	&	1.6	&	-	&	13.5	\\
1502+037	&	0.57	&	0.47	&	1	&	3.4	&	6	&	6	\\
1502+290	&	0.24	&	-	&	-	&	1.5	&	-	&	6.3	\\
1508+081	&	0.08	&	-	&	-	&	2.8	&	-	&	35	\\
1510+157	&	0.33	&	0.88	&	1	&	3.8	&	11.6	&	11.6	\\
1511+227	&	0.1	&	-	&	-	&	1	&	-	&	10	\\
1514+333	&	0.17	&	-	&	-	&	1	&	-	&	6.1	\\
1518-036	&	0.45	&	0.84	&	1	&	2.2	&	4.9	&	4.9	\\
1518+154	&	0.19	&	-	&	-	&	0.9	&	-	&	4.8	\\
1526+377	&	0.44	&	-	&	-	&	5	&	-	&	11.4	\\
1527+352	&	0.43	&	-	&	-	&	3.4	&	-	&	8	\\
1529+243	&	0.06	&	-	&	-	&	1.9	&	-	&	31.2	\\
1529+357	&	0.06	&	-	&	-	&	1.2	&	-	&	20.4	\\
1529+110	&	0.11	&	-	&	-	&	0.7	&	-	&	6.2	\\
1530+155	&	0.44	&	0.39	&	0.86	&	2.8	&	5.4	&	6.3	\\
1534+139	&	0.24	&	-	&	-	&	2.6	&	-	&	11	\\
1538+147	&	0.31	&	0.83	&	1	&	1.6	&	5.3	&	5.3	\\
1538+011	&	0.49	&	0.52	&	1	&	3.7	&	7.4	&	7.4	\\
1539+118	&	0.24	&	-	&	-	&	2	&	-	&	8.5	\\
1542+354	&	0.28	&	-	&	-	&	2.8	&	-	&	9.9	\\
1543+373	&	0.35	&	-	&	-	&	2.1	&	-	&	6.1	\\
1544+082	&	0.2	&	-	&	-	&	1.1	&	-	&	5.3	\\
1547+386	&	0.64	&	-	&	-	&	4.7	&	-	&	7.4	\\
1548+216	&	0.14	&	-	&	-	&	3.7	&	-	&	26.1	\\
1548+053	&	0.17	&	0.73	&	0.47	&	5.3	&	14.5	&	30.8	\\
1549+261	&	0.07	&	-	&	-	&	0.4	&	-	&	6.1	\\
1549+202	&	0.44	&	0.93	&	1	&	10.6	&	24	&	24	\\
1555+357	&	0.16	&	-	&	-	&	1.4	&	-	&	8.8	\\
1557+187	&	0.33	&	-	&	-	&	2.4	&	-	&	7.3	\\
1557+020	&	0.07	&	-	&	-	&	0.4	&	-	&	5.2	\\
1559+158	&	0.28	&	-	&	-	&	2.7	&	-	&	9.7	\\
1600+021	&	0.07	&	-	&	-	&	4.7	&	-	&	66.8	\\
1601+017	&	0.07	&	-	&	-	&	1.4	&	-	&	20.1	\\
1602+375	&	0.44	&	-	&	-	&	2.9	&	-	&	6.6	\\
1605+412	&	0.44	&	-	&	-	&	3.3	&	-	&	7.5	\\
1610+225	&	0.15	&	-	&	-	&	2.4	&	-	&	15.9	\\
1619+128	&	0.09	&	-	&	-	&	0.8	&	-	&	8.8	\\
1621+251	&	0.17	&	-	&	-	&	2.7	&	-	&	15.8	\\
1622+128	&	0.09	&	-	&	-	&	1	&	-	&	11.2	\\
1622+239	&	0.06	&	-	&	-	&	0.9	&	-	&	15.2	\\
1622+123	&	0.16	&	0.58	&	0.38	&	1.1	&	2.6	&	6.9	\\
1623+271	&	0.08	&	-	&	-	&	0.9	&	-	&	11.3	\\
1625+214	&	0.56	&	-	&	-	&	4.7	&	-	&	8.5	\\
1625+278	&	0.1	&	-	&	-	&	1.8	&	-	&	18	\\
1626+150	&	0.21	&	-	&	-	&	1.1	&	-	&	5.1	\\
1626+397	&	0.1	&	-	&	-	&	8.4	&	-	&	84.4	\\
1634+270	&	0.2	&	-	&	-	&	2.1	&	-	&	10.5	\\
1635-033	&	0.56	&	0.75	&	1	&	3.9	&	7	&	7	\\
1635+159	&	0.13	&	-	&	-	&	1.4	&	-	&	10.5	\\
1635+112	&	0.15	&	-	&	-	&	1.5	&	-	&	10.1	\\
1641+300	&	0.15	&	-	&	-	&	0.9	&	-	&	5.9	\\
1642+134	&	0.17	&	-	&	-	&	3.3	&	-	&	19.5	\\
1651+271	&	0.45	&	-	&	-	&	3.1	&	-	&	6.9	\\
1656+124	&	0.1	&	-	&	-	&	2.5	&	-	&	25.4	\\
1657+120	&	0.06	&	-	&	-	&	1	&	-	&	16.2	\\
1658+067	&	0.14	&	-	&	-	&	1.3	&	-	&	9.1	\\
1658+285	&	0.17	&	-	&	-	&	1	&	-	&	5.6	\\
1659+300	&	0.09	&	-	&	-	&	1.1	&	-	&	12.4	\\
1702+298	&	0.1	&	-	&	-	&	0.8	&	-	&	8.1	\\
1707+229	&	0.13	&	-	&	-	&	0.8	&	-	&	6	\\
1708+078	&	0.18	&	-	&	-	&	5.6	&	-	&	30.9	\\
1711+281	&	0.34	&	-	&	-	&	2.1	&	-	&	6.1	\\
1716+316	&	0.38	&	-	&	-	&	2	&	-	&	5.3	\\
1717+229	&	0.25	&	0.9	&	0.88	&	1.4	&	5	&	5.6	\\
1722+343	&	0.06	&	-	&	-	&	0.4	&	-	&	6.8	\\
1733+035	&	0.41	&	0.94	&	1	&	2.1	&	5.2	&	5.2	\\
1736+114	&	0.2	&	-	&	-	&	5.8	&	-	&	29.1	\\
1737+315	&	0.22	&	-	&	-	&	1.3	&	-	&	5.7	\\
1739+311	&	0.13	&	-	&	-	&	0.7	&	-	&	5.4	\\
1739+185	&	0.52	&	-	&	-	&	4.5	&	-	&	8.6	\\
1746+093	&	0.55	&	-	&	-	&	12.9	&	-	&	23.5	\\
1746-015	&	0.32	&	0.91	&	1	&	21.4	&	67.2	&	67.2	\\
1746+160	&	0.07	&	-	&	-	&	0.4	&	-	&	5.2	\\
1750+270	&	0.56	&	-	&	-	&	3.6	&	-	&	6.5	\\
1751+064	&	0.19	&	-	&	-	&	14.8	&	-	&	77.6	\\
1756+135	&	0.39	&	0.33	&	0.67	&	3.7	&	6.3	&	9.4	\\
1756+034	&	0.21	&	-	&	-	&	7.1	&	-	&	33.9	\\
1757+237	&	0.11	&	-	&	-	&	1.3	&	-	&	11.8	\\
1757+012	&	0.29	&	0.4	&	0.5	&	8.5	&	14.7	&	29.6	\\
1757+018	&	0.22	&	0.4	&	0.41	&	2.8	&	5.4	&	13.1	\\
1758-024	&	0.47	&	-	&	-	&	7.6	&	-	&	16.1	\\
1759+139	&	0.33	&	0.52	&	0.85	&	4.2	&	10.8	&	12.7	\\
1803+110	&	0.23	&	-	&	-	&	4.9	&	-	&	21.4	\\
1805+099	&	0.1	&	-	&	-	&	1.3	&	-	&	13.1	\\
1806+109	&	0.23	&	-	&	-	&	4.9	&	-	&	21.1	\\
1808+099	&	0.11	&	-	&	-	&	1.6	&	-	&	14.7	\\
1810+265	&	0.11	&	-	&	-	&	1.1	&	-	&	10.4	\\
1811-022	&	0.3	&	0.81	&	1	&	15.6	&	51	&	51	\\
1812+014	&	0.42	&	0.32	&	0.64	&	17.4	&	26.6	&	41.8	\\
1812+031	&	0.09	&	-	&	-	&	12.1	&	-	&	134.6	\\
1829+147	&	0.1	&	-	&	-	&	0.9	&	-	&	8.7	\\
1831+123	&	0.16	&	-	&	-	&	1.5	&	-	&	9.1	\\
1833+327	&	0.4	&	-	&	-	&	14.5	&	-	&	36.3	\\
1836+171	&	0.1	&	-	&	-	&	3.3	&	-	&	32.5	\\
1837+048	&	0.14	&	-	&	-	&	0.9	&	-	&	6.8	\\
1838+098	&	0.48	&	-	&	-	&	11.6	&	-	&	24.1	\\
1838+133	&	0.25	&	-	&	-	&	7.4	&	-	&	29.7	\\
1841-062	&	0.1	&	-	&	-	&	14.5	&	-	&	144.9	\\
1843+002	&	0.15	&	-	&	-	&	4.8	&	-	&	31.8	\\
1844-053	&	0.31	&	-	&	-	&	37.8	&	-	&	121.8	\\
1845-045	&	0.1	&	-	&	-	&	7	&	-	&	70.1	\\
1846+264	&	0.17	&	-	&	-	&	0.8	&	-	&	4.9	\\
1846-030	&	0.33	&	-	&	-	&	16.9	&	-	&	51.2	\\
1847+074	&	0.08	&	-	&	-	&	0.6	&	-	&	8	\\
1848-061	&	0.14	&	-	&	-	&	3.7	&	-	&	26.1	\\
1848+350	&	0.16	&	0.5	&	1	&	1.2	&	7.3	&	7.3	\\
1848+033	&	0.06	&	-	&	-	&	1.6	&	-	&	27.3	\\
1856+173	&	0.14	&	-	&	-	&	1.1	&	-	&	7.5	\\
1857+008	&	0.41	&	-	&	-	&	18	&	-	&	43.9	\\
1858+127	&	0.16	&	-	&	-	&	4.2	&	-	&	26.3	\\
1901+299	&	0.2	&	-	&	-	&	3.8	&	-	&	19	\\
1901+319	&	0.11	&	-	&	-	&	1.6	&	-	&	14.1	\\
1905+102	&	0.16	&	-	&	-	&	15.7	&	-	&	98	\\
1912+016	&	0.37	&	-	&	-	&	21.1	&	-	&	57	\\
1912+140	&	0.32	&	-	&	-	&	47.6	&	-	&	148.8	\\
1918+393	&	0.11	&	-	&	-	&	2.1	&	-	&	19.3	\\
1921+191	&	0.08	&	-	&	-	&	1.7	&	-	&	21.2	\\
1922+196	&	0.18	&	-	&	-	&	3.2	&	-	&	17.9	\\
1925+054	&	0.13	&	-	&	-	&	3.5	&	-	&	27.2	\\
1926+151	&	0.07	&	-	&	-	&	8.1	&	-	&	115.6	\\
1929+155	&	0.06	&	-	&	-	&	0.3	&	-	&	5.5	\\
1930+190	&	0.06	&	-	&	-	&	5.4	&	-	&	89.8	\\
1931+178	&	0.08	&	-	&	-	&	1.3	&	-	&	16	\\
1933+012	&	0.25	&	-	&	-	&	1.9	&	-	&	7.5	\\
1933+201	&	0.06	&	-	&	-	&	0.8	&	-	&	13.7	\\
1933+166	&	0.09	&	-	&	-	&	4	&	-	&	44.5	\\
1935+185	&	0.32	&	-	&	-	&	2.9	&	-	&	8.9	\\
1938+212	&	0.29	&	-	&	-	&	1.6	&	-	&	5.4	\\
1946+024	&	0.15	&	-	&	-	&	1	&	-	&	6.5	\\
1948+237	&	0.13	&	-	&	-	&	0.7	&	-	&	5.1	\\
1950+135	&	0.07	&	-	&	-	&	1.8	&	-	&	25	\\
1950+254	&	0.4	&	-	&	-	&	3.9	&	-	&	9.9	\\
2000+018	&	0.11	&	-	&	-	&	8.8	&	-	&	79.9	\\
2004+326	&	0.05	&	0.93	&	0.43	&	0.4	&	3.3	&	7.6	\\
2010-051	&	0.52	&	-	&	-	&	2.8	&	-	&	5.3	\\
2011+195	&	0.23	&	0.77	&	0.64	&	1.5	&	4.2	&	6.7	\\
2012+234	&	0.08	&	-	&	-	&	9.5	&	-	&	119.1	\\
2012+264	&	0.39	&	-	&	-	&	5.2	&	-	&	13.3	\\
2016+094	&	0.17	&	-	&	-	&	1.3	&	-	&	7.4	\\
2018+296	&	0.06	&	-	&	-	&	3.6	&	-	&	59.6	\\
2018+126	&	0.18	&	-	&	-	&	1.4	&	-	&	7.7	\\
2018+209	&	0.22	&	-	&	-	&	1.1	&	-	&	5.2	\\
2019+095	&	0.08	&	-	&	-	&	1.1	&	-	&	13.3	\\
2019+178	&	0.09	&	-	&	-	&	0.7	&	-	&	7.6	\\
2019+018	&	0.41	&	0.48	&	1	&	2.1	&	5.3	&	5.3	\\
2020+076	&	0.28	&	-	&	-	&	13.4	&	-	&	47.9	\\
2021+206	&	0.07	&	-	&	-	&	0.4	&	-	&	5.4	\\
2022+213	&	0.16	&	-	&	-	&	1.9	&	-	&	11.8	\\
2024+117	&	0.14	&	-	&	-	&	1.9	&	-	&	13.3	\\
2029+189	&	0.08	&	-	&	-	&	1.5	&	-	&	19.3	\\
2030+072	&	0.65	&	0.72	&	1	&	3.7	&	5.6	&	5.6	\\
2030+243	&	0.44	&	-	&	-	&	2.4	&	-	&	5.4	\\
2030+257	&	0.13	&	-	&	-	&	2	&	-	&	15.4	\\
2036-084	&	0.46	&	-	&	-	&	3.1	&	-	&	6.8	\\
2044-029	&	0.56	&	-	&	-	&	11.8	&	-	&	21.1	\\
2045+233	&	0.25	&	-	&	-	&	1.3	&	-	&	5.2	\\
2046+148	&	0.18	&	-	&	-	&	1	&	-	&	5.5	\\
2049+149	&	0.18	&	-	&	-	&	1.3	&	-	&	7.2	\\
2053+228	&	0.11	&	-	&	-	&	1.3	&	-	&	12	\\
2054+083	&	0.28	&	0.7	&	0.74	&	5.6	&	14.8	&	20	\\
2057+117	&	0.61	&	-	&	-	&	6.9	&	-	&	11.4	\\
2100+140	&	0.76	&	-	&	-	&	4.9	&	-	&	6.4	\\
2105+216	&	0.16	&	-	&	-	&	1.3	&	-	&	7.9	\\
2106+143	&	0.54	&	-	&	-	&	3.3	&	-	&	6.2	\\
2111+302	&	0.34	&	-	&	-	&	2.8	&	-	&	8.3	\\
2113+088	&	0.2	&	-	&	-	&	3	&	-	&	14.9	\\
2118+372	&	0.06	&	-	&	-	&	1.1	&	-	&	18.8	\\
2120+406	&	0.08	&	-	&	-	&	0.5	&	-	&	5.9	\\
2120+168	&	0.07	&	-	&	-	&	1.9	&	-	&	26.5	\\
2121+299	&	0.2	&	-	&	-	&	4	&	-	&	20.1	\\
2130+127	&	0.23	&	-	&	-	&	3.4	&	-	&	14.9	\\
2131+381	&	0.31	&	-	&	-	&	6.6	&	-	&	21.4	\\
2132+226	&	0.14	&	-	&	-	&	0.8	&	-	&	5.7	\\
2132-015	&	0.71	&	0.7	&	1	&	4.5	&	6.3	&	6.3	\\
2135+156	&	0.77	&	-	&	-	&	4.3	&	-	&	5.5	\\
2136+329	&	0.4	&	0.88	&	1	&	2	&	5	&	5	\\
2137+172	&	0.81	&	-	&	-	&	8.2	&	-	&	10.1	\\
2140+103	&	0.16	&	-	&	-	&	1.7	&	-	&	10.6	\\
2142+042	&	0.41	&	0.63	&	1	&	2.8	&	6.9	&	6.9	\\
2143+104	&	0.29	&	-	&	-	&	1.4	&	-	&	5	\\
2144+334	&	0.06	&	-	&	-	&	0.5	&	-	&	8.3	\\
2145+151	&	0.07	&	-	&	-	&	1.7	&	-	&	24	\\
2148+143	&	0.34	&	0.49	&	0.76	&	10.7	&	23.7	&	31	\\
2149+328	&	0.09	&	-	&	-	&	0.9	&	-	&	9.5	\\
2149+211	&	0.13	&	-	&	-	&	1	&	-	&	7.8	\\
2149+215	&	0.14	&	-	&	-	&	0.7	&	-	&	4.7	\\
2152+144	&	0.31	&	0.47	&	0.62	&	4.2	&	8.6	&	13.9	\\
2153+124	&	0.36	&	-	&	-	&	3.7	&	-	&	10.2	\\
2154+372	&	0.09	&	-	&	-	&	1.3	&	-	&	14.1	\\
2156+297	&	0.33	&	-	&	-	&	3.6	&	-	&	11	\\
2200+119	&	0.64	&	0.36	&	1	&	10.5	&	16.4	&	16.4	\\
2203+239	&	0.11	&	-	&	-	&	1.6	&	-	&	14.7	\\
2203+292	&	0.09	&	-	&	-	&	2.1	&	-	&	23.8	\\
2207+374	&	0.24	&	-	&	-	&	2.6	&	-	&	11	\\
2209+082	&	0.3	&	-	&	-	&	7.5	&	-	&	25.1	\\
2218+413	&	0.33	&	-	&	-	&	3.4	&	-	&	10.3	\\
2219-048	&	0.56	&	0.41	&	1	&	3.3	&	5.8	&	5.8	\\
2222+306	&	0.09	&	-	&	-	&	0.6	&	-	&	7.1	\\
2222+049	&	0.21	&	-	&	-	&	5	&	-	&	24	\\
2222+216	&	0.58	&	-	&	-	&	3.5	&	-	&	6	\\
2227+261	&	0.22	&	-	&	-	&	1.4	&	-	&	6.3	\\
2227+248	&	0.28	&	-	&	-	&	1.6	&	-	&	5.7	\\
2230+267	&	0.1	&	-	&	-	&	0.6	&	-	&	6.1	\\
2231+359	&	0.2	&	-	&	-	&	3.6	&	-	&	17.9	\\
2231+410	&	0.26	&	0.87	&	0.77	&	1.5	&	4.5	&	5.8	\\
2233-065	&	0.39	&	-	&	-	&	1.9	&	-	&	4.9	\\
2234+340	&	0.17	&	-	&	-	&	2.1	&	-	&	12.4	\\
2234+334	&	0.25	&	-	&	-	&	1.7	&	-	&	6.7	\\
2238-012	&	0.51	&	-	&	-	&	2.6	&	-	&	5.1	\\
2239+334	&	0.14	&	-	&	-	&	1.5	&	-	&	10.4	\\
2243+315	&	0.06	&	-	&	-	&	0.8	&	-	&	12.7	\\
2244+368	&	0.08	&	-	&	-	&	1.7	&	-	&	20.8	\\
2245+356	&	0.18	&	-	&	-	&	1.1	&	-	&	5.9	\\
2246+183	&	0.84	&	0.5	&	1	&	4.8	&	5.6	&	5.6	\\
2247+113	&	0.15	&	-	&	-	&	3	&	-	&	19.8	\\
2247+134	&	0.49	&	-	&	-	&	9.1	&	-	&	18.5	\\
2248+222	&	0.41	&	0.57	&	0.94	&	2.1	&	4.8	&	5.1	\\
2250+035	&	0.52	&	0.52	&	1	&	2.6	&	5	&	5	\\
2250+379	&	0.17	&	-	&	-	&	4.3	&	-	&	25.1	\\
2251+159	&	0.37	&	0.31	&	0.58	&	4.6	&	7.2	&	12.5	\\
2253-083	&	0.33	&	-	&	-	&	1.6	&	-	&	4.8	\\
2257+314	&	0.12	&	-	&	-	&	0.6	&	-	&	4.9	\\
2302+224	&	0.16	&	-	&	-	&	1.3	&	-	&	8.2	\\
2309+093	&	0.32	&	0.73	&	0.88	&	8.3	&	22.9	&	26.1	\\
2310+051	&	0.07	&	-	&	-	&	2.1	&	-	&	30.4	\\
2327+031	&	0.21	&	-	&	-	&	1.3	&	-	&	6.1	\\
2329+296	&	0.42	&	0.46	&	0.91	&	4.2	&	9.2	&	10	\\
2331+399	&	0.59	&	-	&	-	&	3.9	&	-	&	6.6	\\
2334+153	&	0.84	&	0.42	&	1	&	4.5	&	5.4	&	5.4	\\
2334+048	&	0.2	&	-	&	-	&	1.2	&	-	&	6.2	\\
2335+136	&	0.34	&	0.92	&	1	&	1.7	&	5	&	5	\\
2335+128	&	0.79	&	0.3	&	1	&	4.8	&	6.1	&	6.1	\\
2336+381	&	0.22	&	-	&	-	&	1.1	&	-	&	4.8	\\
2338+039	&	0.54	&	0.49	&	1	&	13	&	24.2	&	24.2	\\
2339+260	&	0.12	&	-	&	-	&	1.4	&	-	&	11.4	\\
2342+292	&	0.24	&	-	&	-	&	1.1	&	-	&	4.8	\\
2348-026	&	0.45	&	-	&	-	&	2.8	&	-	&	6.2	\\
2349+217	&	0.42	&	-	&	-	&	2	&	-	&	4.8	\\
2349+290	&	0.32	&	-	&	-	&	2.5	&	-	&	7.9	\\
2351+339	&	0.12	&	-	&	-	&	0.6	&	-	&	4.9	\\
2351+400	&	0.45	&	-	&	-	&	4.8	&	-	&	10.6	\\
2353+283	&	0.34	&	-	&	-	&	2.2	&	-	&	6.5	\\
2354+144	&	0.34	&	0.46	&	0.73	&	2.8	&	5.9	&	8.1	\\
2355+313	&	0.13	&	-	&	-	&	1.2	&	-	&	9.1	\\
2358+236	&	0.32	&	-	&	-	&	1.5	&	-	&	4.7	\\
2359+108	&	0.44	&	0.27	&	0.66	&	2.6	&	3.9	&	5.9	\\
\end{supertabular}

%\label{lastpage}
\end{document}